\documentstyle[aps,pre,epsf,eqsecnum]{revtex}
\begin{document}
\draft
\title
{
\bf\Large
L\'{e}vy flights in quenched random force fields
}
\author{Hans C. Fogedby}
\address{
\thanks{Permanent address}
Institute of Physics and Astronomy,
University of Aarhus, DK-8000, Aarhus C, Denmark\\
and\\
NORDITA, Blegdamsvej 17, DK-2100, Copenhagen {\O}, Denmark
}
\date{\today}
\maketitle
\begin{abstract}
L\'{e}vy flights, characterized by the microscopic step
index $f$, are for $f<2$ (the case of rare events)
considered in short range and long
range quenched random force fields with arbitrary vector character
to 
first loop order
in an expansion about the critical dimension $2f-2$
in the short range case
and 
the critical
fall-off exponent $2f-2$ in the long range case.
By means of a dynamic renormalization group
analysis based on the momentum shell integration method,
we determine flows, fixed point, and the associated
scaling properties for the probability distribution
and the frequency and wave number dependent diffusion
coefficient. Unlike the case of ordinary Brownian motion
in a quenched force field characterized by a single critical 
dimension or fall-off exponent $d=2$, two critical dimensions appear in the 
L\'{e}vy case. A critical dimension (or fall-off exponent) $d=f$ below which
the diffusion coefficient exhibits anomalous scaling behavior,
i.e, algebraic spatial behavior and long time tails,
and a critical dimension (or fall-off exponent) $d=2f-2$ below which the
force correlations characterized by a non trivial
fixed point become relevant. 
As a general result
we find in all cases that the dynamic exponent $z$,
characterizing the mean square displacement, locks onto the
L\'{e}vy index $f$, {\em independent} 
of dimension and
{\em independent}
of the presence of weak quenched disorder.
\end{abstract}
\section{Introduction}
There is a current interest in the dynamics of fluctuating 
manifolds
in quenched random environments \cite{fisher93}. 
This fundamental issue in
modern condensed matter physics is encountered in 
problems as diverse as
vortex motion in high temperature superconductors, 
moving interfaces in
porous media, and random field magnets and spin glasses.
In this context the
simplest case is that of a random walker in a random 
environment,
corresponding to a zero dimensional fluctuating manifold. 
This
problem has been treated extensively in the literature 
\cite{alex81,havlin87,bouch90}
and many results are known.

In the case of ordinary Brownian motion, characterized by a finite mean
square step, in a pure environment without disorder, the 
central limit
theorem \cite{feller71} implies that the statistics of the 
walk
is given by a Gaussian
distribution with a mean square deviation proportional to the 
number
of steps or, equivalently, the elapsed time, i.e., the mean 
square
displacement is
\begin{equation}
\langle r^2(t)\rangle\,\propto\, Dt^{2/z},
\label{msd}
\end{equation}
where the dynamic exponent $z$ assumes the value $z=2$ for Brownian walk;
$D$ is 
the effective
diffusion coefficient for the process and $\langle\cdots\rangle$ denotes
an ensemble average.

There are, however, many interesting processes in nature 
which are
characterized by {\em anomalous diffusion} with dynamic exponent $z\neq 2$, 
owing to the
statistical properties of the environments \cite{havlin87,bouch90}.
Examples are found in chaotic systems \cite{geisel85}, turbulence
\cite{gross85,bohr93}, flow in fractal geometries \cite{sahimi87}, and
L\'{e}vy flights \cite{shles85,fogedby92}; these cases generally lead
to enhanced diffusion or superdiffusion with $z<2$; we note that
the ballistic case corresponds to $z=1$. The other case of
subdiffusion or dispersive behavior with $z>2$ is encountered
in various constrained systems such as doped crystals, glasses or
fractals \cite{scher74,shles74,alex82,blumen84,blumen86}.

Independent of the spatial dimension $d$, ordinary 
Brownian
motion traces out a manifold of fractal dimension $d_F=2$
\cite{hughes81}.
In the
presence of a quenched disordered force field in $d$ dimensions the
Brownian walk is unaffected for $d>d_F$, i.e., for $d$ larger than
the critical 
dimension $d_{c1}=d_F$ the walk is transparent and the
dynamic exponent
$z$ locks onto the value $2$ for the pure 
case.
Below the critical dimension $d_{c1}=2$ the long time characteristics
of the walk is changed to subdiffusive behavior with $z>2$
\cite{fisher84,derrida83,luck83}. In $d=1$, 
$\langle r^2(t)\rangle\propto [\log{t}]^4$,
independent of the strength of the quenched disorder \cite{sinai82}.

L\'{e}vy flights constitute an interesting generalization of ordinary
Brownian walks.
Here the step size is drawn from a L\'{e}vy distribution
characterized by the step index $f$ \cite{feller71,fogedby92}. 
The L\'{e}vy distribution
has a long range algebraic tail corresponding to large but infrequent
steps, so-called {\em rare events}. This step distribution has the interesting
property that the central limit theorem does not hold in its usual
form. For $f>2$ the second moment or mean square deviation of the step
distribution is finite, the central limit theorem holds, and the dynamic
exponent $z$ for the L\'{e}vy walk locks onto $2$, corresponding to ordinary
diffusive behavior; however, for $f<2$ the mean square step deviation
diverges, the rare large step events prevail determining the
long time behavior, and the dynamic exponent $z$ depends on
the microscopic step index $f$ according to the relationship 
$z=f~~(f<2)$, indicating
anomalous enhanced diffusion, that is superdiffusion
\cite{fogedby92,klafter90,blumen89}. The ``built in'' superdiffusive
character of L\'{e}vy flights has been used to model a variety
of physical processes such as
self diffusion in micelle systems \cite{ott90} and transport
in heterogeneous rocks \cite{klafter90}.

In a recent letter \cite{fogedby94} we considered L\'{e}vy flights in 
the presence of a
quenched isotropic random force field and examined the interplay between the
``built in'' superdiffusive behavior of the L\'{e}vy flights and the
pinning effect of the random environment generally leading to
subdiffusive behavior. Generalizing the discussion in
\cite{fisher84,luck83,aronovitz84} we found that in the case of
enhanced diffusion for $f<2$ the dynamic exponent $z$ locks onto
$f$, {\em independent}
of the presence of weak disorder. On the other hand, we could still
identify a critical dimension $d_{c1}=2f-2$, depending on the step
index $f$ for $f<2$.
Below $d_{c1}$ the quenched
disorder becomes {\em relevant} as indicated by the emergence of a
non-trivial
fixed point in the renormalization group analysis. 
We also briefly discussed the behavior of the subleading diffusive term.
Here another critical dimension, $d_{c2}=f$ enters in the sense that 
for dimensions less than $d_{c2}$ the correction to the ordinary
diffusion coefficient diverges. This feature has also been discussed
from a heuristic point of view in ref. \cite{bouch88}.
Note that in the Brownian case for $f=2$ the critical
dimensions coincide, i.e., $d_{c1}=d_{c2}$.

In the present paper we discuss in more detail L\'{e}vy flights in
quenched random force field with arbitrary range and vector character thus
corroborating and extending the results obtained in ref. \cite{fogedby94}.  
The paper is
organized in the following manner. In Sec. II we introduce and discuss
L\'{e}vy flights in a pure environment.  Next we derive the Langevin
equation for L\'{e}vy flights in a force field and the associated Fokker
Planck equation for the probability distribution.  In Sec. III we 
establish perturbation theory for 
the probability distribution in the Fokker Planck equation averaged over
the quenched random force field and set up
the renormalization group analysis.
In order to extract the scaling
properties and control divergent contributions we dilute the degrees of
freedom by means of the momentum shell integration method.  Finally we
derive differential renormalization group equations to first loop order
for the parameters in the model.  In Sec. IV we determine flows and
fixed point structure
for the renormalization group equations in some detail in the
case of short and long range isotropic force fields and determine the
scaling properties. Since the discussion for anisotropic 
force fields is quite analogous we summarize this case in the Appendix.
We conclude the paper with a summary and a conclusion in Sec V.
A range of technical issues are for the record deferred to an
extensive Appendix.
\section{The Model}
It is our aim to analyze the scaling properties of L\'{e}vy flights in a
quenched random force field.  For that purpose let us first discuss and
summarize the properties of L\'{e}vy flights in a pure environment.

\subsection{L\'{e}vy flights}

We consider a particle or for that matter a gas of non-interacting
particles performing independent isotropic random motion in $d$
dimensions.  For the step size distribution we assume a normalized
L\'{e}vy
distribution
\cite{feller71,fogedby92,hughes81,levy37,nature,shles82,shles86}
\begin{equation}
p(\bbox{\eta})d^d\eta = \frac{f \eta_0^f}{S_d} \eta^{-1-f} d\eta
d\Omega~~~~~~~\text{and}~~~~~~~ S_d=\frac{2\pi^{\frac{d}{2}}}{(d/2-1)!}~.
\label{ld}
\end{equation}
Here $\bbox{\eta}$ is the microscopic step and the algebraic
behavior in the large step regime is characterized by the microscopic
step index $f$.  In order to ensure normalization of the distribution
$p(\bbox{\eta})$ we introduce a lower cut-off $\eta_0$ of the
order of a microscopic length and assume a positive step index, $f>0$. 
$S_d$ is the full solid angle on the surface of a unit sphere in $d$ 
dimensions.
In Fig. 1 we have depicted the L\'{e}vy distribution
$P(\bbox{\eta})$.

The macroscopic physics ensuing from the L\'{e}vy distribution for the
microscopic elementary step depends entirely on the range characteristics
of $P(\bbox{\eta})$.  For $f>2$ the second moment or mean square
step, 
$
\langle\eta^2\rangle= 
\int p(\bbox{\eta})\eta^2d^d\eta = \frac{f}{f-2} \eta^2_0 ~ ,
$
exists  and a characteristic step size is given by the root mean square
deviation $\langle\eta^2\rangle^{\frac{1}{2}}$. For $1< f < 2$ the
second moment diverges, however, the mean step
$
\langle \eta \rangle = \int p(\bbox{\eta}) \eta d^d{\eta} =
\frac{f}{f-1} \eta_0 
$
is finite, defining an effective step size.  In the interval $0<f<1$ the
first moment diverges and even a mean step size is not defined.

For a L\'{e}vy flight consisting of $N$ steps the end-to-end distance of
the random walk is 
$\bbox{r}_N=\sum_{i=1}^N\bbox{\eta}_i$ or, assuming one step per
unit time, $t = N$, 
$
\bbox{r}(t) = \sum_{i=1}^{t}\bbox{\eta}_i ~ ,
$
and it follows generally that the mean square displacement for statistically
independent step events is given by
\begin{equation}
\langle r^2(t)\rangle=t\langle\eta^2\rangle~.
\label{msd2}
\end{equation}
Consequently, for general step distributions with a finite second moment
or mean square step, as in particular the case for L\'{e}vy distribution 
for $f>2$,
the mean square displacement is proportional to the elapsed time.  This
behavior is characteristic of ordinary Brownian motion leading to
diffusive behavior for a gas of Brownian walker and corresponds to the
dynamic exponent $z = 2$ in Eq. (\ref{msd}) \cite{feller71,pathria72}. 

On the other hand, for L\'{e}vy flight with step index 
$f<2$ the second moment $\langle
\eta^2\rangle$ diverges, Eq.(\ref{msd2}) is undefined and
we cannot define a mean
square displacement and a dynamic exponent $z$ according to the heuristic
definition in Eq. (\ref{msd}). However, this point will be clarified below. 
in terms of the probability distribution. 

The probability distribution for the random walker is given by
$P(\bbox{r},t)=\langle\delta(\bbox{r}-\bbox{r}(t))\rangle$ and is easily
derived by means of the method of characteristic functions 
\cite{feller71,hughes81,shles82}.  We obtain, noting that $t$ is discrete,

\begin{equation}
P(\bbox{r},t) = 
\int e^{i\bbox{k\cdot r}}\left[p(\bbox{k})\right]^t \frac{d^d k}{(2\pi)^d}~,
\label{pd}
\end{equation}
where $p(\bbox{k}) $ is the Fourier transform of the step distribution 
$p(\bbox{\eta})$,
$
p(\bbox{k}) = \int e^{-i\bbox{k\cdot\eta}}p(\bbox{\eta})d^d\eta ~.
$
At large $\bbox{r}$ Eq. (\ref{pd}) samples the small $\bbox{k}$ region.  
For small $\bbox{k}$ the behavior of $p(\bbox{k})$ is controlled 
by the algebraic power law tail of $p(\bbox{\eta})$.  We find 
\begin{equation}
p(\bbox{k}) = 1 - D(k\eta_0)^{\mu}\simeq \exp{[-D \eta_0^{\mu} k^{\mu}]} ~,
\label{plt}
\end{equation}
where $D$ is a dimensionless geometrical factor and $\mu$ is a scaling index.  
For $f > 2$ the second moment $\langle \eta^2 \rangle$ exists and $\mu$ locks 
onto 2; 
for $f<2, ~\langle\eta^2\rangle$ diverges and $\mu = f$.  By insertion we 
obtain
\begin{equation}
P(\bbox{r},t)=\int
\exp{[i\bbox{k\cdot r} -\mid t\mid D\eta_0^\mu k^\mu]}
\frac{d^dk}{(2\pi)^d}~.
\label{pd1}
\end{equation}
In Fig. 2 we have shown the dependence of the scaling index $\mu$ on the 
step index $f$.
From the structure of Eq. (\ref{pd1}) we infer that $P(\bbox{r},t)$ has the 
scaling form 
\begin{equation}
P(\bbox{r},t)=
\mid t\mid^{-\frac{d}{\mu}}G(r/\mid t\mid^{\frac{1}{\mu}})~.
\label{sf}
\end{equation}
For $f>2, \mu = 2$ and the scaling function $G(x)$ takes a Gaussian form 
characteristic of ordinary Brownian motion, 
$G(x) \simeq \exp(\text{const.} x^2)$.  
This is a consequence of the central limit theorem \cite{feller71} which 
here implies a universal behavior characterized by the exponent $\mu = 2$.  
For $f<2, \mu=f$ and the scaling function $G(x)$ can in general not be given 
explicitly in terms of known functions.  For $\mu =1$, the ballistic case 
$r \sim t$, we find the Cauchy distribution 
$G(x)\simeq (1 + \text{const.}x^2)^{-((d+1)/2)}$.  It is, moreover, easy to 
show that $G(x) \rightarrow\text{const.}$ for $x \rightarrow 0$ and 
$G(x) \rightarrow 0$ for $x \rightarrow \infty$.

The scaling properties of L\'{e}vy flights are described by Eqs. (\ref{pd1}) 
and (\ref{sf}).  From Eq. (\ref{sf}) we infer the mean square displacement 
$\langle r^2 (t) \rangle \sim t^{2/\mu}$, i.e., according to Eq. (\ref{msd}) 
the dynamic exponent $z = \mu$, indicating superdiffusive behavior.  
However, as mentioned above this reasoning is not correct for L\'{e}vy flights.
The expression in Eq. (\ref{msd2}) is not defined for L\'{e}vy flights since 
$\langle \eta^2 \rangle$ diverges and the scaling regime must be defined 
with some care if we wish to give meaning to Eq. (\ref{msd}).  One way is 
to confine the L\'{e}vy flights to a volume of linear size $L$; the mean 
square displacement is then given by 
$
\langle r^2 (t) \rangle_L = \int_r r^{2}P (\bbox{r},t) d^d r
$
,
where
$
V = L^d.
$
Using the scaling form in Eq. (\ref{sf}) with 
\begin{equation}
G(r) = \left(\frac{1}{D^{1/\mu}\eta_0}\right)^d 
\int \exp[i\bbox{k\cdot r}/(D^{\frac{1}{\mu}} \eta_0) - 
k^{\mu}] \frac{d^d k}{(2\pi)^d}~,
\end{equation}
we obtain, changing variables, the mean square displacement
\begin{equation}
\langle r^2 (t) \rangle_L = 
|t| ^{2/\mu} \int_{(L/t^{1/\mu})^d} r^2 G(r) d^d r~.
\label{msd1}
\end{equation}
From Eq. (\ref{sf}) and (\ref{msd1}) we infer that the characteristic 
time in the problem is the time $ \sim L^{\mu} $ it takes the random walker 
to traverse the volume $V \sim L^d$.  At long times, i.e., 
for $t^{1/\mu} \gg L$,
\begin{equation}
\langle r^2 (t) \rangle_L = |t|^{2/\mu} \int_0 r^2 G(r)d^dr  = 
\text{const.}|t|^{2/\mu}
\end{equation}
and comparing with Eq. (\ref{msd}) we deduce  the dynamic exponent $z = \mu$.  
At short times, $t^{1/\mu}\ll L$, using $G(r) \sim r^{- \mu - d}$ for $r$ 
large, we obtain 
\begin{equation}
\langle r^2 (t) \rangle_L=\text{const.}|t| L^{2 - \mu} ~,
\end{equation}
in accordance with Eq. (\ref{msd2}).
For L\'{e}vy flights with 
$\mu <2~ \langle r^2(t)\rangle_L$ diverges for $t \rightarrow \infty$.  
In the Brownian case $\mu = 2$ and $\langle r^2 (t)\rangle \propto |t|$ in 
both cases in accordance with Eq. (\ref{msd2}).  Summarizing, in order to 
characterize anomalous superdiffusion arising from L\'{e}vy flights by 
means of the mean square displacement we must conceptually confine the 
flights to a box of size $L$ and define the scaling region for 
$t \gg L^{\mu}$.  Alternatively, we can discuss the scaling properties and 
the identification of $z$ by means of the scaling form in Eq. (\ref{sf}) 
where the argument $r$ controls the range in question.
\subsection{Langevin equation for L\'{e}vy flights}
It is convenient to discuss L\'{e}vy flights in terms of a Langevin equation 
with ``power law'' noise [32,33].  In an arbitrary force field 
$\bbox{F}(\bbox{r})$, representing the quenched disordered environment, 
the equation takes the form
\begin{equation}
\frac{d\bbox{r}(t)}{dt}=\bbox{F}(\bbox{r}(t))+\bbox{\eta}(t) ~.
\label{le}
\end{equation}
Here $\bbox{\eta}(t)$ is the instantly correlated power law white noise 
with the isotropic distribution given in Eq. (\ref{ld}) at a particular 
time instant.

The microscopic steps $\bbox{\eta}_i$ with distribution $p(\bbox{\eta}_i)$ 
constituting a L\'{e}vy flight are discrete processes, so properly 
speaking the Langevin equation (\ref{le}) is the continuum limit of the 
corresponding difference equation defined for a discrete time step $\Delta$.  
This limit is singular and it is instructive to discuss the limiting 
procedure in some detail.  

Limiting our discussion to the force-free case, $\bbox{F} = \bbox{0}$, 
setting $\bbox{r}_n =\bbox{r}(t_n), \bbox{\eta}_n = \bbox{\eta}(t_n)$, and 
$t_n = n\Delta$, the difference equation is given by
$
(\bbox{r}_{n+1} - \bbox{r}_n)/\Delta = \bbox{\eta}_n
$
with solution $\bbox{r}_n = \Delta \sum_{p=0}^n \bbox{\eta}_p$.  From the 
definition of 
$P(\bbox{r},n)=\langle\delta(\bbox{r}-\bbox{r}_n)\rangle$, 
we obtain, 
using $\delta(\bbox{r})=\int\exp{(-i\bbox{k\cdot r})}d^d k/(2\pi)^d$ 
together with
$\exp{(i\bbox{k\cdot r}_n)} = 
\prod_{p=0}^n\exp{(i\Delta\bbox{k\cdot \eta}_p)}$ 
for $P(\bbox{k},n)=\int\exp(i\bbox{k\cdot r})P(\bbox{r},n)d^dr$,
\begin{equation}
P(\bbox{k},n) = \langle\exp{(i\Delta\bbox{k}\cdot\bbox{\eta})}\rangle^n
\label{dst}
\end{equation}
and the issue is to define the continuum limit 
$\Delta \rightarrow 0$, keeping $t=n \Delta$ fixed,\  
of the expression for $P(\bbox{k},n)$ in Eq. (\ref{dst}).
For simplicity we first consider the case of a Gaussian distribution 
for $p(\bbox{\eta})$, i.e., the case of ordinary Brownian motion, 
$p(\mbox{\boldmath$\eta$}) = (1/2 \sigma\pi^{1/2})^d \exp(- \eta^2/4 \sigma^2)$ 
of width 2$\sigma$. Evaluating $\langle\exp(i\Delta\bbox{k\cdot\eta} )\rangle$ 
by completing the square in the exponent we obtain 
$P(\bbox{k}, n) = \exp{(-\Delta^2\sigma^2 k^2 n)}$. In the continuum limit 
$\Delta\sigma^2\rightarrow D$ 
and we have the Gaussian distribution
\begin{equation}
P(\bbox{k}, t) = \exp{[-D k^2 t]}~.
\end{equation}
Correspondingly, the white noise condition, i.e., 
uncorrelated noise, 
$\langle\eta^{\alpha}_n \eta^{\beta}_m\rangle = 
\sigma^2 \delta_{nm} \delta_{\alpha \beta}$, becomes in the continuum limit
\begin{equation}
\langle\eta^{\alpha}(t)\eta^{\beta} (t')\rangle = 
D\delta^{\alpha\beta}\delta(t - t')~.
\end{equation}
Consequently, in the Gaussian case we must in the continuum limit 
$\Delta \rightarrow 0$ scale the width $\sigma$ of the noise distribution 
to infinity in order to obtain a finite diffusion coefficient 
$D = \sigma^2 \Delta$.
In the L\'{e}vy case the noise distribution $p(\bbox{\eta})$ is given by 
Eq. (\ref{ld}) and from Eq.(\ref{plt}), 
$\langle\exp(i\Delta\bbox{k\cdot\eta})\rangle = 
1-D(k\Delta\eta_0)^f$ for $f<2$ 
and $k\Delta\eta_0\ll 1$, i.e., $P(\bbox{k},n) = [1 -D(k\Delta\eta_0)^f]^n$.  
Using the relation, $(1 - x/n)^n \rightarrow \exp(-x)$ for 
$n \rightarrow \infty$, 
we obtain in the continuum limit 
$\Delta \rightarrow 0$ and $t = n \Delta$ fixed,
\begin{equation}
P(\bbox{k},t)=\exp{[-D\eta^f_0\Delta^{f-1}k^f t]}~.
\end{equation}
In order to eliminate the discrete time step $\Delta$ and keep the 
coefficient $D$ 
fixed we must renormalize the cut off $\eta_0$ in the L\'{e}vy distribution 
(\ref{ld}) according to $\eta_0^f \Delta^{f - 1} = 1$.  We notice that the 
border line case is $f=1$.  For $1<f<2 ~\eta_0 \rightarrow \infty$ for 
$\Delta \rightarrow 0$, i.e., the cut-off moves out to infinity.  
For $0<f<1$, on the other hand, $\eta_0 \rightarrow 0$ for 
$\Delta \rightarrow 0$.  We also notice that $\Delta \eta_0 \rightarrow 0$ 
so that $k \Delta \eta_0\ll 1$ is satisfied for all $f$.

Summarizing, we can without loss of generality discuss L\'{e}vy flights 
(and Brownian motion) in terms of a continuous Langevin equation provided we 
scale the underlying discrete noise distributions accordingly.  Note, however, 
that these renormalizations are not observable, the problem at hand is defined 
by the Langevin equations (\ref{le}).

L\'{e}vy flights in an arbitrary force field are in principle described 
by the Langevin equation (\ref{le}) together with the distribution in 
Eq. (\ref{ld}) for the noise $\mbox{\boldmath$\eta$}$.  For a given force 
field the only random aspect resides in the noise $\mbox{\boldmath$\eta$}$ 
which drives the position $\bbox{r}$ of the random walker; the force field 
$\bbox{F}(\bbox{r})$ acts as a static background.  However, for a random 
force field modelling the quenched static environment the Langevin equation 
(\ref{le}) harbours two different kinds of stochasticity and it is convenient 
to recast the problem in terms of an associated Fokker-Planck equation, 
thereby absorbing the fluctuating L\'{e}vy noise $\bbox{\eta}(t)$.
\subsection{Fokker-Planck equation for L\'{e}vy flights}
In the absence of the force field, i.e., $\bbox{F}(\bbox{r}) = \bbox{0}$, it 
follows from Eq. (\ref{pd1}) that $P(\bbox{k},t)$ satisfies the equation
\begin{equation}
\frac{\partial P(\bbox{k},t)}{\partial t} = - D_1 k^\mu P(\bbox{k},t)~,
\label{fp1}
\end{equation}
where we have absorbed the renormalized cut-off $\eta_0$ in the L\'{e}vy 
coefficient $D_1 = D\eta^{\mu}_0$.  The form of Eq. (\ref{fp1}) leads us to 
suggest the following Fokker-Planck equation for a L\'{e}vy walker in a 
force field:
\begin{eqnarray}
\frac{\partial P(\bbox{r}, t)}{\partial t} = 
- \bbox{\nabla}\cdot(\bbox{F}(\bbox{r})P(\bbox{r},t))
+ D_1 \nabla^{\mu} P(\bbox{r},t) + D_2 \nabla^2 P (\bbox{r},t)~.
\label{fp2}
\end{eqnarray}
Here the first term on the right hand side of Eq. (\ref{fp2}) is the usual 
drift term due to the motion of the walker in the force field, the second 
term arises from Eq. (\ref{fp1}) where 
we have introduced the ``fractional gradient operator'' $\nabla^\mu$ as the 
Fourier transform of $-k^\mu$. ~ $\nabla^\mu \sim r^{-d-\mu}$  is a 
spatially nonlocal integral operator reflecting the long range character of the
L\'{e}vy steps; 
for $\mu = 2$ it reduces to the usual Laplace operator describing ordinary 
diffusion \cite{hcfogedby}.  Finally, we have for later purposes 
included the ordinary diffusion term, $D_2 \nabla^2 P(\bbox{r}, t)$, 
originating from the next to leading long range part of the distribution 
$p(\bbox{\eta})$ and 
corresponding to the next leading term of order $k^2$ in the
expansion in Eq. (\ref{plt}).  
The derivation of the Fokker-Planck equation is discussed in more detail 
in Appendix A.

For the quenched random force field we assume a Gaussian distribution,
\begin{eqnarray}
p(\bbox{F}(\bbox{r}))\propto
\exp{[-\frac{1}{2}
\int d^drd^dr'
F^\alpha({\bf{r}})
\Delta^{\alpha\beta}({\bf{r}}-{\bf{r}^\prime})^{-1}
F^\beta({\bf{r}}')]}~,
\label{fd} 
\end{eqnarray}
with the
spatial correlations given by
\begin{equation} 
\langle F^\alpha(\bbox{r})F^\beta(\bbox{r}')\rangle_F=
\Delta^{\alpha\beta}(\bbox{r}-\bbox{r}')~.
\label{fc} 
\end{equation} 
Here $\Delta^{\alpha \beta} (\bbox{r} - \bbox{r}')$ is the force correlation 
function expressing the range and vector character of $\bbox{F}(\bbox{r})$ 
and $\langle\cdots\rangle_F$ denotes an average over the force field 
according to the distribution (\ref{fd}).

In the unconstrained case $\Delta^{\alpha\beta}$ is diagonal; 
however, generally the force field breaks up into a longitudinal curl-free 
part $\bbox{E}$ and a transverse divergence-free part $\bbox{B}$, 
i.e., $\bbox{F} = \bbox{E}+\bbox{B}$, where 
$\bbox{\nabla}\cdot\bbox{B} = 0$ 
and $\bbox{\nabla}\times\bbox{E} =\bbox{0}$ 
\cite{bouch90,aronovitz84,fisher85}.  
Assuming that the cross correlation of $\bbox{E}$ and $\bbox{B}$ vanishes, 
$\langle E^\alpha B^\beta \rangle = 0$, we have in Fourier space
\begin{eqnarray}
\langle E^\alpha(\bbox{k})E^\beta(\bbox{k}^\prime)\rangle_F =&& 
(2\pi)^d\delta(\bbox{k}+
\bbox{k}^\prime)\left[\frac{k^\alpha k^\beta}{k^2}\right]\Delta^L(\bbox{k})
\label{fc1}
\\
\langle B^\alpha(\bbox{k}) B^\beta (\bbox{k}')\rangle_F =&& 
(2\pi)^d\delta(\bbox{k}+\bbox{k}')\left[\delta^{\alpha\beta} - 
\frac{k^\alpha k^\beta}{k^2}\right]\Delta^T (\bbox{k})~.
\label{fc2}
\end{eqnarray}
The case of an unconstrained force field then corresponds to 
$\Delta^L = \Delta^T = \Delta$ and we obtain the force correlation function
\begin{equation} 
\langle F^\alpha(\bbox{k})F^\beta (\bbox{k}')\rangle_F=
(2\pi)^d\delta(\bbox{k}+\bbox{k}')
\delta^{\alpha\beta}\Delta(\bbox{k})~.
\label{fc3}  
\end{equation}
The range of the force correlations is characterized by the functions
$\Delta^L(\bbox{k})$ and $\Delta^T(\bbox{k})$.

In the case of a {\it finite range}, e.g., 
$\Delta^{L,T}(\bbox{r}) \propto \exp (-rk_0) \left / r^{d - 2} \right.$, 
with range parameter $1/k_0$ we have 
$\Delta^{L,T} (\bbox{k}) \propto 1/(k^2 + k^2_0)$.  
In the long wavelength limit 
$k \rightarrow 0$ ~ $\Delta^{L,T}(\bbox{k})\rightarrow\text{const}.$, 
corresponding to the {\it zero range} case 
$\Delta^{L,T} (\bbox{r}) \propto \delta(\bbox{r})$.  For {\it infinite range} 
the force correlations are characterized by the exponent $a$, 
$\Delta^{L,T} (\bbox{r})\propto r^{-a}, \Delta^{L,T}(\bbox{k})\propto k^{a-d}$.

Summarizing, we characterize a general force field by the following 
expressions for the longitudinal and transverse force correlation functions:
\begin{eqnarray}
\Delta^L (\bbox{k})=&& \Delta^L_1 + \Delta_2^L k^{a_L - d}
\label{fcf1}
\\
\Delta^T (\bbox{k}) =&& \Delta_1^T + \Delta^T_2 k^{a_T - d}~.
\label{fcf2}
\end{eqnarray}
We have here for completion introduced two separate 
indices, $a_L$ and $a_T$, for the 
power law behavior of the longitudinal and transverse correlations, 
respectively.  For 
$a_{L,T} > d, \Delta^{L,T} (\bbox{k}) \rightarrow \Delta_1^{L,T}$, 
in the long wavelength limit and we effectively retrieve the zero range case; 
for 
$a_{L,T} < d, \Delta^{L,T}(\bbox{k})\rightarrow\Delta^{L,T}_2 k^{a_{L,T} - d}$ 
and the specific long range behavior enters in the 
$\bbox{k} \rightarrow\bbox{0}$ limit.
\section{Renormalization group theory}
The problem of analyzing the asymptotic long time scaling properties of 
L\'{e}vy flights in a random quenched force field has now been reduced to an 
analysis of the random Fokker-Planck equation  (\ref{fp2}), in conjunction 
with the force distribution in Eq. (\ref{fd}) and the force 
correlation functions in Eqs. (\ref{fc}-\ref{fc2}).  
In particular we 
wish to evaluate the scaling behavior of the distribution $P(\bbox{r}, t)$ 
averaged with respect to the force field, i.e., 
$\langle P(\bbox{r}, t)\rangle_F$.
\subsection{Perturbation theory}
There are a variety of techniques available in order to treat the 
random Fokker-Planck equation (\ref{fp2}).  Applying the 
Martin-Siggia-Rose formalism in functional form 
\cite{martin73,domi78,cde78,bausch76,peliti85} and using either the 
replica method \cite{bouch90} or an explicit causal time dependence 
\cite{fisher84,cde78,peliti85}, one can average over the quenched force 
field and construct an effective field theory.  A more direct method, 
which we shall adhere to in the present discussion, amounts to an 
expansion of the Fokker-Planck equation (\ref{fp2}) in powers of 
the force field and an average over products of $\bbox{F}(\bbox{r})$ 
according to the distribution in Eq. (\ref{fd}).

Defining the Laplace-Fourier transform
\begin{equation}
P(\bbox{k},\omega)=\int d^3 rdt \exp(i\omega t-i\bbox{k\cdot r})
P(\bbox{r},t)\theta(t)~,
\label{lft} 
\end{equation}
where $\theta(t)$ is the step function, we obtain, introducing for later 
purposes the dimensionless coupling strengths $\lambda_L$ and $\lambda_T$ 
for the vertices coupling to $\bbox{E}$ and $\bbox{B}$, respectively, the 
Fokker-Planck equation
\begin{equation}
P(\bbox{k},\omega)=G_0(\bbox{k},\omega)P_0 (\bbox{k})
+G_0 (\bbox{k},\omega)\int\frac{d^d p}{(2\pi)^d}
[(-i)\lambda_L\bbox{k\cdot E}(\bbox{k}-\bbox{p})
+(-i)\lambda_T\bbox{k\cdot B}(\bbox{k}-\bbox{p}]P(\bbox{p},\omega)
\label{fopl}
\end{equation}
Here the force field is averaged according to Eqs. (\ref{fc1}-\ref{fc2}) for
all pairwise contractions according to the Wick's theorem following from
the Gaussian distribution in Eq. (\ref{fd}),
$P_0 (\bbox{k}) = P(\bbox{k},t=0)$
is the initial distribution, and we have, moreover, introduced
the unperturbed propagator or
Green's function
\begin{equation}
G_0 (\bbox{k},\omega)=\frac{1}{-i\omega + D_1 k^\mu + D_2 k^2} ~.
\end{equation}
The integral equation (\ref{fopl}) immediately lends 
itself to a direct expansion in powers of 
$\bbox{E}, \bbox{B}$ or $\lambda_L, \lambda_T$.
In order to discuss the various perturbative contributions it is 
convenient to represent Eq. (\ref{fopl}) diagrammatically 
as done in Fig. 3
\cite{bouch90,aronovitz84,ma75,ma76,forster77,fogedby78}.
Here
$P(\bbox{k}, \omega)$ is characterized by a solid bar, 
$P_0(\bbox{k})$ by a cross, the 
vertices $- i \lambda_L$ and $ - i \lambda_T$ by dots, the force fields 
$\bbox{E}$ and $\bbox{B}$ by wiggly lines, and $G_0(\bbox{k},\omega)$ by 
a directed arrow.

Hence, iterating Eq. (\ref{fopl}) in powers of $\lambda_L$ and $\lambda_T$ and 
averaging over the force fields, keeping one component fixed, we 
identify perturbative corrections to 1) the self energy, 2) the vertex 
functions, and 3) the force correlation functions.  Defining the self 
energy $\Sigma(\bbox{k}, \omega)$ by means of the Dyson equation,
\begin{equation}
G (\bbox{k},\omega) = G_0(\bbox{k},\omega)+
G_0(\bbox{k},\omega)\Sigma(\bbox{k},\omega)G(\bbox{k},\omega) ~,
\label{de}
\end{equation}
shown diagrammatically in Fig. 4, where the renormalized propagator is 
indicated by a solid directed line and the self energy by a circle, 
we derive the renormalized  Fokker-Planck equation shown in Fig. 5,
\begin{equation}
P(\bbox{k},\omega)=G(\bbox{k},\omega)P_0(\bbox{k})
+G(\bbox{k},\omega)\int\frac{d^d p}{(2 \pi)^d}
\left[(-i)\bbox{\Lambda}_L(\bbox{k},\bbox{p},\omega)
\cdot\bbox{E}(\bbox{k}-\bbox{p})
+(-i)\bbox{\Lambda}_T(\bbox{k},\bbox{p},\omega)
\cdot\bbox{B}(\bbox{k}-\bbox{p})\right]
P(\bbox{p},\omega) ~.
\end{equation}
Here $\bbox{\Lambda}_L(\bbox{k},\bbox{p},\omega)$ and 
$\bbox{\Lambda}_T(\bbox{k},\bbox{p},\omega)$, depicted as circles in 
Fig. 5, are the 
renormalized vertex functions; to lowest order 
$\bbox{\Lambda}_{L,T}(\bbox{k},\bbox{p},\omega) = 
\lambda_{L,T}\bbox{k}$. In a similar manner 
we extract corrections to the force correlation function from the 4-point 
vertex function $\Gamma(\bbox{k},\bbox{p},\bbox{l},\omega)$ 
in Fig. 6 and 
the contraction
$\int\Gamma (\bbox{k},\bbox{k},\bbox{l},\omega)
G(\bbox{k}-\bbox{l},\omega)d^d l/(2\pi)^d$,
also shown in Fig. 6.  To lowest order 
$\Gamma(\bbox{k},\bbox{k},\bbox{l},\omega)=
- k^\alpha\Delta^{\alpha\beta}(l)(k-l)^\beta$, summed over 
$\alpha$ and $\beta$, where $\Delta^{\alpha \beta}$ is the force 
correlation function defined in Eqs. (\ref {fc}-\ref{fc2}).
To second order in $\lambda_L$ and $\lambda_T$ or first order in 
$\Delta^L$ and $\Delta^T$, corresponding to 
first loop order in the field theoretic formulation 
\cite{fisher84,martin73,domi78,cde78,bausch76,peliti85} we find 
diagrammatic contributions to $\Sigma, \bbox{\Lambda}_{L,T}$, and 
$\Gamma$ shown in Fig. 7, Fig. 8, and Fig. 9, respectively.

Let us first discuss the self energy correction.  Solving the Dyson 
equation (\ref{de}) the self energy $\Sigma (\bbox{k},\omega)$ 
enters in the 
renormalized propagator,
\begin{eqnarray}
G(\bbox{k},\omega)=
\frac{1}{-i\omega+D_1 k^\mu+D_2 k^2-\Sigma(\bbox{k},\omega)} ~,
\label{fpr}
\end{eqnarray}
and directly determines the diffusional character of the random walker.
In Appendix B we  discuss in some detail the evaluation of 
$\Sigma (\bbox{k},\omega)$ to leading order in $k^2$ on the basis of 
the diagrams in Fig. 7.

In the case of an isotropic unconstrained zero range force correlation 
function, $\lambda_L = \lambda_T = \lambda$ and 
$\Delta^L(\bbox{k}) = \Delta^T(\bbox{k}) = \Delta$, we have in particular
\begin{equation}
\Sigma(\bbox{k},\omega)=
-\lambda^2\Delta\int\frac{d^d p}{(2\pi)^d}
\bbox{k}\cdot(\bbox{k}/2+\bbox{p})G_0(\bbox{k}/2+\bbox{p},\omega) ~.
\end{equation}
To leading order in $k^2$ the static contribution, $\Sigma(\bbox{k},0)$, 
given 
in Appendix B,
only contributes to the ordinary 
diffusion term $D_2 k^2$ in Eq. (\ref{fpr}); there is no correction to the 
anomalous L\'{e}vy term $D_1k^\mu$. We shall see later that this has a 
profound 
effect on the scaling properties of L\'{e}vy flights.  For the correction to 
the 
diffusion coefficient $D_2$ we then find, performing the integration over 
the solid angle,
\begin{equation}
\delta D_2=\frac{1}{2}\lambda^2\Delta\frac{S_d}{d(2\pi)^d}
\int^\Lambda_0dp\frac{D_1(d-\mu)p^{\mu + d-1}+D_2 (d-2)p^{d+1}}
{[D_1p^\mu + D_2p^2]^2}~,
\label{dc}
\end{equation}
where $S_d$ is given in Eq. (\ref{ld}) and we have introduced a UV 
cut-off corresponding to a microscopic length of order $1/\Lambda$.
In the long wavelength limit $p \rightarrow 0$ the integrand in 
Eq. (\ref{dc}) is dominated by the leading L\'{e}vy term $\sim p^\mu$ and 
simple power counting shows that the integral is convergent for $d >\mu$  
yielding a correction to $D_2$.  For  $d<\mu$ the integral diverges in the 
infrared limit $p \rightarrow 0$ and we need a renormalization group 
approach in order to disentangle the true asymptotic scaling behavior.  
We encounter here the first {\em critical dimension} $d_{c2}=\mu$ 
characterizing
the behavior of the ordinary diffusion coefficient $D_2$.
In the Brownian case \cite{fisher84,derrida83,luck83} $\mu = 2$ and 
the critical dimension is $2$.

In a similar way we can discuss the vertex correction 
$\bbox{\Lambda}(\bbox{k,p},\omega)$ on the basis of the diagrams in Fig. 8 
in the isotropic case.  The vertex function $\bbox{\Lambda}$ describes 
the coupling of the random walker to the quenched force field.  
In Appendix B we discuss in detail the evaluation of $\bbox{\Lambda}$.  
We obtain
\begin{equation}
\bbox{\Lambda}(\bbox{k},\bbox{p},\omega) = 
\lambda\bbox{k}-\lambda^3\Delta
\int\frac{d^dl}{(2\pi)^d} 
(\bbox{k-l})\bbox{k\cdot(p-l)}
G_0(\bbox{k-l},\omega)
G_0(\bbox{p-l},\omega)
\label{vc}
\end{equation}
from which we extract to leading order in $k$ and for $\omega = 0$ the 
perturbative correction to $\lambda$,
\begin{equation}
\delta\lambda = -\lambda^3\Delta\frac{S_d}{d(2\pi)^d}
\int^\Lambda _0 dp \frac{p^{d+1}}{[D_1 p^\mu + D_2 p^2]^2}~.
\label{vc1}
\end{equation}
Similarly to our discussion of the self energy, the integral in 
Eq. (\ref{vc1}) is convergent for $d>2\mu-2$, whereas for $d<2\mu - 2$ 
the correction $\delta \lambda$ diverges in the far infrared limit 
$p \rightarrow 0$.  We note here the appearance of a second
{\em critical dimension} $d_{c1}=2 \mu -2$, 
characterizing the behavior of the 
vertex correction.
In the Brownian case $\mu = 2$ and both the vertex and self energy 
diverges for $d <2$.  We also note that for $\mu = 2$ the first order 
correction to $D_2$ actually vanishes for $d=2$ thus requiring an expansion 
to second order in $\Delta$ (fourth order in $\lambda$) or, equivalently, 
to two-loop order \cite{fisher84}.

Finally, we discuss the correction to the force correlation function 
$\Delta$ extracted from the contraction of the 4-point vertex function 
depicted in Figs. 6 and 9.  From the results in Appendix B we deduce
\begin{eqnarray}
\Gamma (\bbox{k,p,l},\omega) =
&&-\bbox{k\cdot(p-l)}\lambda^2\Delta\nonumber\\
&&+\lambda^4\Delta^2\int\frac{d^d n}{(2\pi)^d}
\bbox{k\cdot(p-n)}\bbox{(k-n)\cdot(p-l)}
G_0(\bbox{k-n},\omega)
G_0(\bbox{p-n},\omega)
 \nonumber\\
&&+\lambda^4\Delta^2\int\frac{d^d n}{(2\pi)^d} 
\bbox{k\cdot(p-l)}\bbox{(k-n)\cdot(p-l+n)}
G_0(\bbox{k-n},\omega)
G_0(\bbox{p-l+n},\omega)
\label{vc2}
\end{eqnarray}
and we obtain, contracting Eq. (\ref{vc2}), to leading order in $k$ and 
for $\omega = 0$ the perturbative correction to $\Delta$,
\begin{eqnarray}
\delta \Delta = \lambda^2 \Delta^2 \frac{S_d}{(2 \pi)^d} 
(1 - \frac{1}{d}) \int^\Lambda_0 dp \frac{p^{d+1}}{[D_1p^\mu + D_2p^2]^2}~.
\label{nc}
\end{eqnarray}
Also here we note that the correction to $\Delta$ diverges for 
$d<d_{c1}$.
In the Brownian case, $\mu = 2$, the critical dimensions coincide,
i.e., $d_{c1} = d_{c2}$.
\subsection{Momentum shell integration}
In order to disentangle the breakdown of primitive perturbation theory and 
deduce the scaling properties of the force averaged distribution 
$\langle P(\bbox{r},t)\rangle_F$ and the mean square displacement
$\langle \langle r^2(t)\rangle \rangle_F$ we carry out a dynamic 
renormalization group analysis, following the momentum shell 
integration
method \cite{aronovitz84,ma75,forster77,fogedby80}.
This approach is a way of systematically diluting 
the short wavelength degrees of freedom, keeping the long wavelength modes 
controlling the asymptotic scaling behavior.  The method derives from 
Wilson's original momentum space procedure \cite{wilson74} applied to 
static critical phenomena \cite{ma76}
and is an implementation of the real space Kadanoff construction
in momentum space \cite{forster77}.
The application of renormalization 
group theory to dynamic phenomena described by Langevin-type equations was 
initiated in the context of dynamical critical 
phenomena in refs. \cite{ma76,frey94}.  The implementation of 
the momentum shell 
integration method was introduced and discussed in refs.
\cite{ma75,ma76,forster77}.

Here we briefly  review the momentum shell integration method in the context 
of the Fokker-Planck equation (\ref{fopl}), which in the case of an 
isotropic force field is expressed in the symbolic form
\begin{eqnarray}
P = G_0 P_0 + \lambda G_0 FP~.
\label{fps}
\end{eqnarray}
Setting the UV cut-off $\Lambda = 1$, dividing  the wave number interval 
$0<k<1$ into a long wavelength regime $0<k<e^{-\ell}, ~\ell>0$, including 
the long distance scaling region, and a short wavelength ``shell'' region 
$e^{-\ell} < k<1$, the idea is now to average out the short wavelength 
degrees of freedom, using the distribution of the force field in 
Eqs. (\ref{fd}-\ref{fc}) in the shell $e^{-\ell}<k<1$, in order to derive
a renormalized Fokker-Planck equation valid for long wavelengths 
$0<k<e^{-\ell}$. 
Projecting the 
Fokker-Planck equation (\ref{fps}) onto the two regions in wave number 
space we obtain the coupled equation
\begin{eqnarray}
P^\prime =&& 
G^\prime_0 P^\prime_0 + \lambda G^\prime_0 
(FP + F^\prime P + FP^\prime +F^\prime P^\prime)
\label{sfp1}
\\
P =&&
G_0P_0 + \lambda G_0 (FP + F^\prime P + FP^\prime + F^\prime P^\prime)~,
\label{sfp2}
\end{eqnarray}
where the prime refers to the shell $e^{-\ell}<k<1$. 
Averaging the ``long wavelength'' Fokker-Planck equation (\ref{sfp2})
with respect to the 
short wavelength degrees of freedom in the shell we have, noting that 
$\langle F^\prime\rangle_{F^\prime} = 0$, 
\begin{equation}
\langle P \rangle_{F^\prime} = 
G_0 P_0  
+\lambda G_0 FP 
+\lambda G_0 F \langle P^\prime \rangle_{F^\prime} 
+ 
\lambda G_0 \langle F^\prime P^\prime \rangle_{F^\prime}~.
\end{equation}
The evaluation of $\langle P^\prime \rangle_{F^\prime }$ and 
$\langle F^\prime P^\prime \rangle_{F^\prime }$ to a given order in 
$\lambda$ is now achieved by expanding the ``short wavelength'' Fokker-Planck 
equation (\ref{sfp1}) and averaging over $F^\prime $.  Using that 
$G_0FG_0^\prime $ and $G_0F^\prime G_0^\prime F^\prime G_0^\prime $ vanish 
in the long wavelength limit, since two small wave numbers cannot add up to 
a wave number in the shell, and defining the force contraction
according to the notation $F_{c_i}^\prime F_{c_j}^\prime =
\langle F^\prime F^\prime\rangle\delta_{ij}$, we obtain
\begin{eqnarray}
\langle P\rangle_{F^\prime }
=&& 
G_0 P_0 
+ \lambda G_0 F\langle P\rangle_{F^\prime} 
+ \lambda^2 G_0 F_c^\prime G_0^\prime F_c^\prime \langle P\rangle_{F^\prime}
+ \lambda^4 G_0F_{c_1}^\prime G_0^\prime F_{c_2}^\prime G_0^\prime 
F_{c_1}^\prime
G_0^\prime F_{c_2}^\prime \langle P\rangle_{F^\prime}
\nonumber
\\
+&&\lambda^4 G_0 F_{c_1}^\prime G_0^\prime F_{c_2}^\prime G_0^\prime 
F_{c_2}^\prime
G_0^\prime F_{c_1}^\prime \langle P\rangle_{F^\prime }
+\lambda^3 G_0 F_c^\prime G_0^\prime F G_0^\prime F_c^\prime
\langle P\rangle_{F^\prime }
+\lambda^4 G_0 F_c^\prime G_0^\prime F G_0^\prime F G_0^\prime F_c^\prime
\langle P\rangle_{F^\prime } 
\label{symexp}
\end{eqnarray}
which is diagrammatically depicted in Fig. 10.

We note that the momentum shell integration method combined with a 
perturbative expansion is essentially a nonlinear procedure leading to a 
more general Fokker-Planck equation involving higher order force fields 
as shown in Fig. 10 where the last diagram corresponds to two force fields 
$F$ coupling to the distribution P.
More importantly for deriving renormalization group equations we identify 
the same self energy, vertex, and force corrections as in primitive 
perturbation theory but now evaluated within the short wavelength shell.  
In this manner the elimination of the short wavelength degrees of freedom 
enters in the Fokker-Planck equation for the remaining long wavelength 
degrees of freedom.
\subsection{Renormalization group equations}
While the
general expression to first loop order for the self energy, vertex, 
and force correlation corrections and the general renormalization
group equations are given in Appendix B, we here present 
a detailed derivation of the renormalization group equations in the 
isotropic short range case.

Disregarding the two-force term in Eq. (\ref{symexp}) which is of higher 
order in $k$ and therefore becomes irrelevant in the long wavelength limit, 
we obtain the renormalized Fokker-Planck equation
\begin{equation}
[-i \omega + D_1 k^{\mu} + (D_2 + \delta D_2) k^2 ] P (\bbox{k}, \omega) = 
P_0 (\bbox{k} )
+(\lambda+\delta\lambda)\int\frac{d^dp}{(2\pi)^d}
(-i)\bbox{k\cdot F(k-p)}P(\bbox{p},\omega)
\label{eq1}
\end{equation}
and force correlation function
\begin{equation}
\langle F^\alpha (\bbox{k})F^\beta(\bbox{p})\rangle_F = 
(\Delta + \delta\Delta)\delta^{\alpha\beta}(2\pi)^d\delta(\bbox{k+p})~,
\label{eq2}
\end{equation}
both valid for wave numbers in the interval
for $0<k,p<e^{-\ell}$.

The first step in deriving renormalization group equations is then to rescale 
the wave number range to $0 <k,p<1$ and in this manner compensate for the 
eliminated degrees of freedom. Rescaling at the same time frequency, 
probability distribution, and force field, according to
\begin{eqnarray}
\bbox{k}^\prime=&&\bbox{k}e^\ell
\label{rs1}
\\
\omega^\prime =&& \omega e^{\alpha(\ell)}
\label{rs2}
\\
P^\prime (\bbox{k}^\prime,\omega^\prime ) =&& 
P(\bbox{k}, \omega )e^{- \alpha(\ell)}
\label{rs3}
\\
\bbox{F}^\prime (\bbox{k}^\prime ) =&& 
\bbox{F}(\bbox{k})e^{- \beta(\ell)}~,
\label{rs4}
\end{eqnarray}
where $\alpha(\ell)$ and $\beta(\ell)$ are to be determined
subsequently, we
thus obtain the renormalized Fokker-Planck equation
\begin{equation}
[-i\omega^\prime+D_1^\prime(\ell)k^{\prime\mu}
+D_2^\prime(\ell)k^{\prime 2}]
P^\prime(\bbox{k}^\prime,\omega^\prime)=
P^\prime _0 (\bbox{k}^\prime)
+
{(-i)}\lambda^\prime(\ell)
\int \frac{d^d p^\prime }{(2\pi)^d} \bbox{k}^\prime
\cdot\bbox{F}^\prime (\bbox{k}^\prime -\bbox{p}^\prime )
P^\prime (\bbox{k}^\prime, \omega^\prime)
\label{rfp}
\end{equation}
and force correlation function
\begin{equation}
\langle F^{\alpha\prime}(\bbox{k}^\prime ) 
F^{\beta\prime}(\bbox{p}^\prime)\rangle_{F^\prime} = 
\Delta^\prime(\ell) 
\delta^{\alpha\beta}(2\pi)^d\delta(\bbox{k}^\prime+\bbox{p}^\prime)~.
\label{rfc}
\end{equation}
for wave numbers $k'$ and $p'$ in the original range $0<k',p'<1$;
note that $P_0(\bbox{k})= P_0^\prime (\bbox{k}^\prime)$.
We have here introduced the scale dependent
parameters 
$D^\prime_1(\ell),D^\prime_2(\ell),\lambda^\prime(\ell)$, 
and $\Delta^\prime (\ell)$ given by  
\begin{eqnarray}
D_1^\prime (\ell)&=&D_1e^{-\ell\mu +\alpha(\ell)} 
\label{r1}
\\
D_2^\prime (\ell)&=&(D_2+\delta D_2(\ell))e^{-2\ell +\alpha(\ell)} 
\label{r2}
\\
\lambda^\prime(\ell)&=&(\lambda + \delta \lambda (\ell)) 
e^{-\ell(1+d)+\alpha (\ell)+ \beta(\ell)} 
\label{r3}
\\
\Delta^\prime(\ell)&=&(\Delta + \delta \Delta(\ell)) e^{d\ell-2 \beta(\ell)}~. 
\label{r4}
\end{eqnarray}
Note that the corrections $\delta D_2(\ell), \delta \lambda (\ell)$, and 
$\delta \Delta (\ell)$ depend on the scale parameter $\ell$ since they are 
evaluated in the short wavelength shell $e^{-\ell} < k < 1$.

In order to allow for an iteration  the renormalization group equations 
are recast in 
differential form by considering an infinitesimal 
scale parameter $\ell$ and expanding the right hand sides of 
Eqs. (\ref{r1}-\ref{r4}).  Defining
\begin{eqnarray}
\alpha (\ell) =&& \int_0^\ell  z(e^\prime) d \ell^\prime
\label{alfa}
\\ 
\beta (\ell) = &&\int^\ell_0 u(\ell^\prime)d\ell^\prime
\label{beta}
\end{eqnarray}
and noting from Eqs. (\ref{dc}), (\ref{vc1}), and (\ref{nc}) 
evaluated on the shell $k,p = 1$ that
\begin{eqnarray}
\delta D_2&\propto&\frac{1}{2}\frac{S_d}{(2\pi)^d}\frac{1}{d}
\frac{D_1(d-\mu)+D_2(d-2)}{[D_1+D_2]^2}\lambda^2\Delta\ell
\label{dcor}
\\
\delta\lambda&\propto&-\frac{S_d}{(2{\bf\pi})^d}\frac{1}{d}
\frac{\lambda^3\Delta}{[D_1 + D_2]^2}\ell
\\
\delta \Delta &\propto&\frac{S_d}{(2\pi)^d}
(1-\frac{1}{d})\frac{\lambda^2 \Delta}{[D_1 + D_2]^2}\ell ~.
\end{eqnarray}
We arrive at the differential renormalization group equations

\begin{eqnarray}
\frac{dD_1}{d\ell} 
&=&(z-\mu)D_1
\label{rg1}
\\
\frac{dD_2}{d\ell} 
&=&(z-2)D_2+A\frac{D_1(d-\mu)+D_2(d-2)}
{[D_1+D_2]^2}\lambda^2 \Delta
\label{rg2}
\\
\frac{d\lambda}{d\ell}&=&(z+u-1-d)\lambda -
2A\frac{\lambda^3\Delta}{[D_1+D_2]^2}
\label{rg3}
\\
\frac{d\Delta}{d\ell} &=& (d-2u)\Delta+2A(d-1)
\frac{\lambda^2 \Delta}{[D_1 + D_2]^2}~, 
\label{rg4}
\end{eqnarray}
where $A = (1/2d)S_d/(2\pi)^d$ is a geometrical factor,
$S_d=2\pi^{d/2}/(d/2-1)!$ (cf. Eq. (\ref{ld})).

In conformity with Ref. \cite{aronovitz84} we have included a vertex 
coupling $\lambda$.  However, since $\lambda^2$ is always associated with 
$\Delta$ in the diagrammatic expansion, the inclusion 
of both $\lambda$ and $\Delta$ is essentially superfluous and we can for 
example set $\lambda = 1$ and discuss the coupling to the force field 
by means of
$\Delta$ alone \cite{fisher84}.  
Thus assuming $d\lambda(\ell)/d\ell = 0$ in Eq. (\ref{rg3}) and solving
for $u(\ell)$ we finally 
obtain the renormalization group equations
\begin{eqnarray}
\frac{dD_1}{d\ell} 
&=&(z -\mu)D_1
\label{rgg1}
\\
\frac{dD_2}{d\ell} 
&=&(z-2)D_2+A\frac{D_1(d-\mu)+D_2(d-2)}
{[D_1 + D_2]^2} \Delta 
\label{rgg2}
\\
\frac{d\Delta}{d\ell} 
&=&(2z-d-2)\Delta-2A(3-d)\frac{\Delta^2}{[D_1+D_2]^2}
\label{rgg3}
\end{eqnarray}
which provides the basis for the discussion of the scaling properties of 
L\'{e}vy flights in an isotropic short range force field.

Deferring details to Appendices B and C we obtain in a precisely analogous 
manner the renormalization group equations for
$D_1$, $D_2$, $\Delta^L_1$, $\Delta^L_2$, $\Delta^T_1$, and $\Delta^T_2$, in 
the case of L\'{e}vy flights in a  general force field with range 
characterized by the indices $a_L$ and $a_T$:
\begin{eqnarray}
\frac{dD_1}{d \ell}=
&&(z-\mu)D_1
\label{rggg1}
\\
\frac{dD_2}{d\ell} =
&&(z-2)D_2 \nonumber\\
&&+A\frac{(\Delta^L_1 + \Delta^L_2) (D_1 (2-\mu - d) + D_2 (-d))}
{[D_1 + D_2] ^2}\nonumber\\
&&+A\frac{(\Delta^T_1 + \Delta^T_2) (D_1 (2d-2) + D_2 (2d-2))}
{[D_1 + D_2]^2}\nonumber\\
&&+A\frac{\Delta^L_2 (d-a_L)(D_1 + D_2)}{[D_1 + D_2]^2}
\label{rggg2}
\\
\frac{d\Delta^L_1}{d\ell} =
&&(2z-d-2)\Delta^L_1-4A\frac{\Delta^L_1+\Delta^L_2}
{[D_1+D_2]^2}\Delta^L_1 \nonumber\\
&&+A(2d-2)\frac{(\Delta^T_1 + \Delta^T_2) (\Delta^L_1 + \Delta^L_2)}
{[D_1+D_2]^2}
\label{rggg3}
\\
\frac{d \Delta^T_1}{d\ell}=
&&(2z-d-2)\Delta^T_1-4A\frac{\Delta^L_1+\Delta^L_2}
{[D_1+D_2]^2}\Delta^T_1\nonumber\\
&&+A(2d-2)\frac{(\Delta^T_1+\Delta^T_2)(\Delta^L_1+\Delta^L_2)}
{[D_1+D_2]^2}
\label{rggg4}
\\
\frac{d\Delta^L_2}{d\ell} =
&&(2z-2-a_L)\Delta^L_2-4A\frac{\Delta^L_1+\Delta^L_2}
{[D_1+D_2]^2}\Delta^L_2
\label{rggg5}
\\
\frac{d\Delta^T_2}{d\ell}=
&& (2z-2-a_T)\Delta^T_2-4A\frac{\Delta^L_1+\Delta^L_2}
{[D_1+D_2]^2}\Delta^T_2
\label{rggg6}
\end{eqnarray}
\section{Discussion}

We now turn to a discussion of the renormalization group 
equations derived in Sec. III. The Eqs. (\ref{rgg1}-\ref{rgg3}) describe the
scaling properties in the isotropic short range case,
whereas
the Eqs. (\ref{rggg1}-\ref{rggg6}) account for the general case of anisotropic
force fields with short or long range correlations. We discuss
the isotropic short range and long range cases in some detail
and summarize the results for the anisotropic cases, deferring details
to the Appendix.

We notice immediately a general feature of the renormalization group
equations (\ref{rggg1}-\ref{rggg6}): The requirement that
the anomalous diffusion coefficient $D_1$, characterizing the
amplitude of the L\'{e}vy term $D_1k^\mu$, stays constant under
renormalization, i.e., $dD_1(\ell)/d\ell = 0$ in Eq. (\ref{rggg1}), 
immediately
implies that the dynamic scale dependent exponent $z(\ell)$ locks onto the
scaling index $\mu$, i.e.,
\begin{eqnarray}
z = \mu~.
\label{dynexp}
\end{eqnarray}
Consequently,
in the case of L\'{e}vy
flights in a weak random force field the long time scaling
behavior is entirely controlled by the leading anomalous L\'{e}vy term
$D_1k^\mu$ and the dynamic exponent $z$ locks onto $\mu$.
In other words, the random force field has no influence on the L\'{e}vy
flights.  The intrinsic long range superdiffusive behavior, that is the
occurrence of {\em rare events}, 
enables the walker to
escape the inhomogeneous pinning environment and the long time behavior
is the same as in the pure case. Below we substantiate this claim
in more detail when we discuss the isotropic short range case.
\subsection{The isotropic short range case}
In the case of an unconstrained short range force field the renormalization 
group equations are given by Eqs. (\ref{rgg1}-\ref{rgg3}).  
They describe how 
the parameters in the renormalized long wavelength Fokker-Planck equation 
(\ref{rfp}) and force correlation function (\ref{rfc}) change as we 
differentiably average out the short wavelength degrees of freedom in the 
shell $e^{-\ell}<k<1$, characterized by the scale parameter $\ell$.
\subsubsection{Renormalization group flow and fixed point structure}
Requiring a constant anomalous diffusion coefficient $D_1$ under
renormalization, i.e., $dD_1(\ell)/d\ell = 0$, Eq. (\ref{rgg1})  
implies that the dynamic scale dependent exponent 
$z(\ell)$ locks onto $\mu$,
$z(\ell)=\mu$, 
and we obtain the renormalization group equations for $D_2$ and $\Delta$,
\begin{eqnarray}
\frac{dD_2}{d\ell} =&& 
(\mu - 2)D_2 + A \frac{D_1(d-\mu) + 
D_2 (d-2)}{[D_1+D_2]^2} \Delta 
\label{RG1}
\\
\frac{d\Delta}{d\ell} =&& 
\epsilon \Delta - 2A (3 - d) \frac{\Delta^2}{[D_1 + D_2]^2} 
\label{RG2}
\\
\epsilon=&&d_{c1}-d
\\
d_{c1} =&& 2 \mu -2 ~.
\label{cdim}
\end{eqnarray}
The Eqs. (\ref{RG1}-\ref{RG2}) determine the renormalization group
flow in the $D_2 - \Delta$ parameter space.
We have introduced the parameter $\epsilon$ and $d_{c1}$
is the critical dimension.
The Eqs. (\ref{RG1}-\ref{RG2}) determine the renormalization group 
flow in the $D_2 - \Delta$ parameter space.

Above the critical dimension for 
$\epsilon < 0$, i.e., $d>d_{c1} = 2 \mu -2$ or $\mu < 1 + d/2$, 
Eqs. (\ref{RG1}-\ref{RG2}) have the trivial Gaussian fixed points 
$D_2^* =0$ and $\Delta^* = 0$, indicating that (i) the subleading diffusion 
term, $D_2k^2$, scales to zero compared with the leading L\'{e}vy term and 
(ii) the quenched disorder, characterized by $\Delta$, scales to zero and 
thus is {\it irrelevant}.  The effective long wavelength 
Fokker-Planck equation takes the form
\begin{eqnarray}
(-i\omega + D_1k^\mu) P(\bbox{k},\omega) = P_0(\bbox{k})
\label{FP}
\end{eqnarray}
and for a particle at the origin at $t = 0$, i.e., $P_0 (\bbox{k}) = 1$, 
we obtain the scaling expressions in Eqs. (\ref{pd1}-\ref{sf}) with 
dynamic exponent $z = \mu$.  Clearly, the physics is characterized by the 
interplay between the dimension $d$ of configuration space and the index $\mu$
specifying the tail of the L\'{e}vy distribution.  For $\mu < 1+d/2$ the 
long range L\'{e}vy steps predominate and control the scaling behavior.

Below the critical dimension for $\epsilon >0$, i.e., $d<d_{c1} = 2 \mu - 2$, 
or $1+ d/2 < \mu <2$, we obtain, solving Eqs. (\ref{RG1}-\ref{RG2}) 
for $dP_2/d\ell=0$ and $d\Delta/d\ell =0$, the non-trivial fixed point 
values for $D_2$ and $\Delta$: 
\begin{eqnarray}
D_2^* &=& - D_1 \frac{(\mu - d)\epsilon}{(2 - \mu )(6-2d)+(2-d)\epsilon}
\label{fixp1}
\\
\Delta^* &=& \epsilon \frac{(D_1 + D_2^*)^2}{2A (3-d)}~.
\label{fixp2}
\end{eqnarray}
The fixed point $D_2^*$ indicates that the subleading diffusive term 
$D_2k^2$ yields a contribution compared to the L\'{e}vy term 
$D_1 k^\mu$.  The fixed point value of the diffusion coefficient $D_2^*$ 
is negative since the pinning environment created by the random force field 
reduces the ordinary diffusion coefficient from its value $D_2^* = 0$ 
for $\mu < 1 + d/2$.  The emergence of the fixed point $\Delta^*$ shows that 
for $d$ less than the critical dimension $d_{c1} = 2 \mu - 2$ the quenched 
disorder in the long wavelength Fokker-Planck equation becomes 
{\it relevant}.  In Fig. 11 we have shown the critical dimension $d_{c1}$ as a 
function of the scaling index $\mu$.  For $\mu = 2$ we have the Brownian 
case $d_{c1}=2$; for $d<\mu<2$ the critical dimension
$d_{c1}$ depends linearly on $\mu$.  Note that 
$d_{c1} =0$ in the ballistic case for $\mu = 1$.  The line 
$d = d_{c2}= \mu$ 
delimits the region $d<d_{c2}$ where naive perturbation theory for 
$\delta D_2$ diverges as discussed in Sec. III.

In the Brownian case $\mu \rightarrow 2$ it follows from Eq. (\ref{fixp1}) 
that $D_2^* \rightarrow - D_1$ so that the L\'{e}vy term $D_1 k^\mu$ 
precisely cancels the diffusive term $D_2k^2$ in the Fokker-Planck 
equation (\ref{fopl}); this is consistent with the fact that there is no 
correction to first loop order or more precisely to first order in 
$\epsilon = d_{c1} - d$ in the Brownian case \cite{fisher84}.  In Fig. 12 we 
have plotted the position of the fixed point $(\Delta^* ,D_2^*)$ as a 
function of the scaling index $\mu$ for 
$\epsilon = d_{c1} - d>0$ or $1 + d/2<\mu<2$.
\subsubsection{Scaling properties}
Introducing the notation
$
P(\bbox{k},\omega,D_1,D_2,\Delta)=\langle P(\bbox{k},\omega)\rangle_F ~,
$
the scaling properties of $\langle P(\bbox{k}, \omega)\rangle_F$ are 
determined in the long wavelength limit by noting that 
$\langle P(\bbox{k},\omega)\rangle_F$ 
can equally well be computed from the original Fokker-Planck equation 
(\ref{fopl}) as from the renormalized equation (\ref{rfp}).  
From the explicit scaling definitions in Eqs.(\ref{rs1}-\ref{rs4}) 
we thus obtain the homogeneity relation,
\begin{equation}
P(\bbox{k}, \omega, D_1, D_2, \Delta) = 
e^{\alpha(\ell)}
P(\bbox{k} e^{\ell},\omega e^{\alpha(\ell)},D_1(\ell),D_2(\ell),\Delta(\ell))
\label{hom1}
\end{equation}
which determines how $\langle P(\bbox{k}, \omega)\rangle_F$ varies as we 
average out the short wavelength degrees of freedom parametrized by the 
scale parameter $\ell$.  Similarly, we can derive a homogeneity relation for 
the wave number and frequency dependent diffusion coefficient 
$D_2(\bbox{k}, \omega, D_1, D_2, \Delta)$, defined according to the Dyson 
equation (\ref{de}) and $k^2D_2(\bbox{k},\omega)=-\Sigma(\bbox{k},\omega)$.
From Eq. (\ref{hom1}) we thus obtain
\begin{eqnarray}
D_2 (\bbox{k},\omega,D_1, D_2,\Delta)
= e^{2\ell-\alpha(\ell)}
D_2(\bbox{k}e^{\ell},\omega e^{\alpha(\ell)},D_1(\ell),D_2(\ell),\Delta(\ell))~.
\label{hom2}
\end{eqnarray}
In the vicinity of the fixed point $(\Delta^*, D^*)$, i.e., for large 
$\ell$, we have, setting from Eqs. (\ref{alfa}) and (\ref{dynexp}) 
$\alpha(\ell) = \mu \ell$ and choosing  wave numbers 
$\bbox{k}$ such that
$\bbox{k}e^{\ell}\sim 1$, 
the scaling forms
\begin{eqnarray}
P(\bbox{k},\omega,D_1,D_2,\Delta)
&=&k^{-\mu}L(k/\omega^{1/\mu},D_1,D_2^*,\Delta^*)
\\
D_2(\bbox{k},\omega,D_1,D_2,\Delta)
&=&k^{\mu - 2}M(k/\omega^{1/\mu},D_1,D_2^*,\Delta^*)~,
\end{eqnarray}
where $L$ and $M$ are scaling functions.  Making use of Eq. (\ref{lft}) 
we also have
\begin{equation}\
P(\bbox{r}, t, D_1, D_2, \Delta) = 
|t|^{-d/\mu} G(r/|t|^{1/\mu}, D_1, D_2^*, \Delta^*) ~,
\label{scal3}
\end{equation}
similar to the scaling form in Eq. (\ref{sf}) in the absence of the force 
field and we infer as in Sec. II a dynamic exponent $z$ equal to the 
L\'{e}vy scaling index $\mu$, i.e., $z = \mu$.

Above the critical dimension, i.e., for $\epsilon <0$ or 
$d>d_{c1} = 2 \mu - 2$, we have the Gaussian fixed point 
$(D_2^*, \Delta^*) = (0,0)$ and we infer from Eq. (\ref{FP}) 
the complex scaling function
\begin{equation}
L(x,D_1, 0,0) = \frac{ix^\mu}{iD_1x^\mu + 1}~.
\label{scalf}
\end{equation}
%
\subsubsection{Scaling relations and long time tails}
Using the matching procedure \cite{aronovitz84,forster77,fogedby80} we can 
also derive scaling relations for the wave number frequency-dependent 
diffusion coefficient $D_2(\bbox{k}, \omega)$. Expanding the renormalization 
group equations (\ref{RG1}) and (\ref{RG2}) to first order in $\epsilon$ 
about the fixed point $(\Delta^*, D_2^*)$ and defining 
$\Delta(\ell) = \Delta^* + \delta \Delta(\ell)$ and 
$D_2(\ell) = D_2^* + \delta D_2(\ell)$ the linearized equations take the 
form 
$d\delta D_2/d\ell=-(2-\mu)\delta D_2+A((d-\mu)/D_1)\delta\Delta$
and
$d\delta\Delta/d\ell=-|\epsilon|\delta\Delta$
with solutions
\begin{eqnarray}
\delta D_2 (\ell) 
&=& \left[ \delta D_2^0 + A \frac{\delta \Delta^0}{D_1} \right] 
e^{-(2-\mu)\ell} - A\frac{\delta \Delta^0}{D_1} 
e^{-|\epsilon|\ell} 
\label{sol1}
\\
\delta \Delta (\ell) &=& \delta \Delta^0 e^{-|\epsilon|\ell} \ .
\label{sol2}
\end{eqnarray}
Here $\delta \Delta^0 = \delta \Delta (0)$ and 
$\delta D_2^0 = \delta D_2 (0)$ and Eqs. (\ref{sol1}-\ref{sol2}) 
hold both above and below the critical dimension $d_{c1}$. Since $z(\ell)$ is 
locked on to $\mu$ and suppressing the dependence on $D_1$ we obtain from 
Eq. (\ref{hom2})
\begin{equation}
D_2 (\bbox{k},\omega,D_2,\Delta) = 
e^{(2-\mu)\ell}
D_2(\bbox{k}e^{\ell},\omega 
e^{\mu\ell},D_2^*+\delta D_2(\ell),\Delta^*+\delta\Delta(\ell)) \ ,
\label{hom22}
\end{equation}
where $\delta D_2(\ell)$ and $\delta \Delta (\ell)$ are given by 
Eqs. (\ref{sol1}-\ref{sol2}).

In the long wavelength limit $\bbox{k} \rightarrow 0$ 
and choosing 
$\ell$ such that $\omega e^{\mu\ell}\simeq 1$ and from 
Appendix B, $D_2(\bbox{0},1,D_2,\Delta)\propto\Delta$, we obtain, inserting 
Eq. (\ref{sol2}), the scaling expression
\begin{equation}
D_2 (\bbox{0},\omega,D_2,\Delta)\propto
\omega^{-\frac{(2-\mu)}{\mu}}
\left[\Delta^* +\delta\Delta^0\omega^{|\epsilon|/\mu} \right]~.
\label{hom3}
\end{equation}
Above the critical dimension, i.e., for $d >d_{c1} = 2 \mu - 2$ or 
$\epsilon < 0, \Delta^* = 0$ and we have
\begin{equation}
D_2 (\bbox{0},\omega,D_2,\Delta)\propto\delta\Delta^0
\omega^{(d-d_{c2})/d_{c2}} ~,
\label{hom9}
\end{equation}
where we have introduced the other critical dimension
\begin{equation}
d_{c2}=\mu ~.
\end{equation}
In the low frequency limit $\omega \rightarrow 0$ the coefficient 
$D_2$ vanishes for 
$d>d_{c2}$ and diverges for $d_{c1} < d < d_{c2}$.  
This behavior is consistent 
with our remarks in Sec. III and clarifies the nature of the divergence.  In 
fact, $d_{c2} = \mu$ plays the role of a second critical dimension controlling 
the behavior of the subleading diffusion coefficient $D_2$ and supports the 
heuristic argument by Bouchaud et al. in \cite{bouch88}.  In the Brownian case 
$\mu = 2$ the two critical dimensions coincide.

Below the critical dimension, $d<d_{c1}$ or $\epsilon>0, \Delta^* >0$ and we 
have the leading behavior

\begin{equation}
D_2 (\bbox{0}, \omega, D_2, \Delta)\propto\Delta^* 
\omega^{- (2-d_{c2})/d_{c2}}
\label{hom5}
\end{equation}
which shows divergent behavior in the low frequency limit.

From $D_2(t) = \int \exp(-i\omega t) D_2 (\omega) d\omega /2\pi$ we 
finally obtain the algebraic long time tails for the time dependent 
diffusion coefficient:
\begin{eqnarray}
D_2 (\bbox{0},t,D_2,\Delta)
&\propto& t^{-d/d_{c2}} 
~~~~\mbox{for}~ d>d_{c1} ~,
\label{hom11}
\\
D_2 (\bbox{0},t,D_2, \Delta) 
&\propto& t^{- d_{c1}/d_{c2}} 
~~\mbox{for}~ d<d_{c1} .
\label{hom7}
\end{eqnarray}
In a similar way we can extract the behavior of 
$D_2 (\bbox{k}, \omega, D_2, \Delta)$ in the long wavelength limit.  
Setting $\omega = 0$ in Eq. (\ref{hom22}) 
and choosing $\ell$ such that $\bbox{k}e^{\ell} \simeq 1$ we obtain
\begin{equation}
D_2 (\bbox{k}, 0, D_2, \Delta)
\propto k^{- (2-\mu)} \left[\Delta^* + \delta \Delta^0 k^{|\epsilon|}\right]~,
\label{hom4}
\end{equation}
i.e.,
\begin{eqnarray}
D_2 (\bbox{k}, 0, D_2, \Delta) 
&\propto& \delta \Delta^0 k^{d - d_{c2}} ~~
~~~~\text{for}~ d>d_{c1} 
\label{hom10}
\\
D_2 (\bbox{k}, 0, D_2, \Delta) 
&\propto& \Delta^* k^{- (2-d_{c2})}~~
~~\mbox{for}~ d<d_{c1}~. 
\label{hom6}
\end{eqnarray}
For $d>d_{c2}~~~D_2$ vanishes in the long wavelength limit 
$\bbox{k} \rightarrow 0$; for $d<d_{c2}~~~D_2$ is divergent, in accordance 
with our previous discussion.

The spatial dependence is inferred from 
$D_2 (\bbox{r}) = \int \exp(i\bbox{kr})D_2 (\bbox{k}) d^d k/(2 \pi)^d$ 
and we derive the algebraic long range fall-off 
\begin{eqnarray}
D_2 (\bbox{r}, 0, D_2, \Delta) 
&\propto& r^{d_{c2} - 2d}~~~~~\mbox{for}~~~ d>d_{c1} 
\label{hom12}
\\
D_2 (\bbox{r}, 0, D_2, \Delta) 
&\propto& r^{2-d-d_{2c}} ~~~\mbox{for}~~~ d<d_{c1} 
\label{hom8}
\end{eqnarray}
In Fig. 11 the line  $d= d_{c2}$, the second critical dimension,
delimits the regions for the behavior of $D_2$.  
For $d>d_{c2} ~~D_2$ converges,
for $d<d_{c2}~~D_2$ is divergent.

It is instructive to consider the renormalization group flow in the
$\Delta - D_2$ plane about a fixed point in more detail. 
This discussion is carried out in Appendix D. Another issue, in the analysis
of the renormalization group equations (3.38)-(3.40) we 
chose to keep the L\'{e}vy coefficient $D_1$ fixed under a renormalization 
group transformation. This requirement leads, among other results, to 
$z = \mu$ which is one of the main conclusions of the present work.  
Clearly, keeping $D_1$ fixed is an arbitrary choice and our scaling results 
cannot depend on this choice.  This point is discussed in Appendix E.
\subsection{The isotropic long range case}
We now turn to a discussion of the case of L\'{e}vy flights in an isotropic
long range random force field characterized by a fall-off exponent $a$.  
The case of Brownian motion in an algebraic long range field has been 
discussed by several authors 
\cite{bouch90,peliti85,bouch88,honk88,honkonen,honk89,honk91} 
both to first and second loop order.  The main conclusion here is that 
provided the force field falls off slowly enough the long range force 
correlations interfere with the Brownian walk and give rise to anomalous 
diffusion in any dimension. For comparison we have summarized
the Brownian case in Appendix F.

Keeping as usual $D_1$ fixed 
by locking $z$ onto $\mu$ we extract from the general equations 
(\ref{rggg1}-\ref{rggg6})
the appropriate renormalization group equations for 
$D_2$ and $\Delta_2 = \Delta$, 
\begin{eqnarray}
\frac{dD_2}{d\ell}
&=& 
(\mu-2)D_2+A\frac{D_1(2d-\mu-a)+D_2(2d-2-a)}
{[D_1 + D_2]^2} \Delta
\\
\frac{d\Delta}{d\ell}
&=&
\epsilon\Delta-4A\frac{\Delta^2}{[D_1+D_2]^2} ~,
\end{eqnarray}
where we have introduced the expansion parameter
\begin{equation}
\epsilon = a_{c1} - a~.
\end{equation}
Here $a_{c1} = 2 \mu - 2$ has the same value as the critical 
dimension defined in 
Eq. (\ref{cdim}) for the short range case.  In the present context $a_{c1}$, of 
course, plays the role of a critical fall-off exponent for the 
long range 
force correlations.

The analysis now proceeds precisely as in the short range case.  
For $\epsilon < 0,$ i.e., $a> a_{c1} = 2 \mu - 2$ or $\mu < 1+ a/2$, we obtain 
the trivial Gaussian fixed points $D_2^* = 0$, and $\Delta^* = 0$, showing 
that (i) the subleading diffusion term $D_2 k^2$ scales to zero compared 
with the leading L\'{e}vy term and (ii) the quenched disorder, characterized 
by $\Delta$, scales to zero and thus is {\it irrelevant}.
The effective long wavelength Fokker-Planck equation takes the form in 
Eq. (\ref{FP}) and we obtain the scaling expressions in 
Eqs. (\ref{pd1}-\ref{sf})
with dynamic exponent $z = \mu$.  In contrast to the short 
range case, where the physics is controlled by the interplay between the 
dimension $d$ of configuration place and the L\'{e}vy  index $\mu$, the 
fall-off exponent $a$ replaces $d$ in the long range case for $a <d$.  
For $a >a_{c1}$ or $\mu<1+a/2$ the long range L\'{e}vy  steps 
predominate and determine the scaling behavior.

For $\epsilon > 0$, i.e., $a < a_{c1}$ or $\mu > 1 + a/2$, we 
obtain the non-trivial fixed point values,
$
D_2^*=-D_1(2d-\mu-a)\epsilon/(4(\mu-2)+(2d-2-a)\epsilon) 
$
and
$
\Delta^* = \epsilon(D_1 + D_2^*)^2/4A
$
,
indicating that the subleading term $D_2k^2$ yields a contribution compared 
to the L\'{e}vy  term $D_1 k^\mu$ and that the quenched disorder becomes 
{\it relevant}.  In the Brownian case 
$\mu \rightarrow 2 ~~~D_2^* \rightarrow - D_2$, i.e., the L\'{e}vy term 
$D_1 k^\mu$ precisely cancels the diffusive term $D_2 k^2$  in the 
Fokker-Planck equation (\ref{fopl}).

To leading order in $\epsilon$ we have the fixed points 
$(D_2^*, \Delta^*) = (0,0)$ for $\epsilon < 0$ and the fixed points,
$
D_2^*=\epsilon[(2d - \mu - a)/4(2-\mu)]D_1 
$
and
$
\Delta^* = \epsilon D_1/4A
$
for $\epsilon > 0$.  Similarly, in the vicinity of either fixed point the 
linearized renormalization group equations
$d\delta D_2/d\ell=-(2-\mu)\delta D_2+A((2d-\mu-a)/D_1)\delta\Delta$
and
$d\delta\Delta/d\ell =-|\epsilon|\delta \Delta$
with solutions of the same form as in Eqs. (\ref{sol1}-\ref{sol2}),
i.e.,
\begin{eqnarray}
\delta D_2 (\ell) 
&=& 
\left[ \delta D_2^0 - A \frac{3 \mu - 2 d - 2}{2 - \mu} 
\frac{\delta \Delta^0}{D_1} \right]
e^{-(2-\mu)\ell} - A \frac{3 \mu - 2d - 2}{2 - \mu} 
\frac{\delta \Delta^0}{D_1}e^{-|\epsilon|\ell}
\label{soll1}
\\
\delta \Delta (\ell)
&=& \delta \Delta^0 e^{-|\epsilon|\ell}~.
\label{soll2}
\end{eqnarray}
For the distribution function $P(\bbox{k}, \omega, D_1, D_2, \Delta)$ and 
diffusion coefficient 
$D_2 (\bbox{k}, \omega, D_1, D_2, \Delta)$ we obtain again for large 
$\ell$ 
and choosing $\bbox{k} e^{\ell} \simeq 1$ the scaling forms Eqs. (\ref{scal1}) 
and (\ref{scal2}).
We also obtain the scaling form 
in Eq. (\ref{scal3}) for $P$,
implying that the dynamic exponent $z$ locks onto $\mu$.

As in the short range case the main conclusion is that in the case of 
L\'{e}vy flights in a weak isotropic long range force field, the long time 
scaling behavior is entirely controlled by the leading anomalous L\'{e}vy 
term $D_1k^\mu$. 
The random force field does not influence the L\'{e}vy flights.  The long 
range superdiffusive behavior enables the walker to escape the 
inhomogeneous  pinning environment and the long time behavior is the same 
as in the pure case.

For $\epsilon < 0,$ i.e., $d > a>a_{c1}$, we have the Gaussian fixed 
point $(D_2^*, \Delta^*) = (0,0)$ and we obtain in particular the complex 
scaling function in Eq. (\ref{scalf}) for $P(\bbox{k}, \omega)$.
Implementing the matching procedure as in the short range case, 
we obtain to lowest order
in $\Delta$ Eqs. (\ref{hom3}) and (\ref{hom4})
For $\epsilon < 0$ or $d>a>a_{c1}~~~~\Delta^* = 0$ and we have
\begin{eqnarray}
D_2(\bbox{0},\omega, D_2, \Delta) 
&\propto& \delta \Delta^0  \omega^{(a-a_{c2})/a_{c2}} 
\\
D_2 (\bbox{k}, 0, D_2, \Delta) 
&\propto & \delta \Delta_0 k^{a - a_{c2}} \ ,
\end{eqnarray}
where $a_{c2}=d_{c2}=\mu$ is the second critical fall-off exponent.
In the low frequency limit $\omega \rightarrow 0$ or long wavelength limit 
$\bbox{k} \rightarrow 0$ $D_2$ vanishes for $a >a_{c2}$ and diverges for 
$a < a_{c2}$.  This behavior is in accordance with the naive perturbation 
theory discussed in Sec. III and in Appendix B; introducing a force 
correlation $\Delta \sim p^{a-d}$ in the integrand in Eq. (\ref{dc}) for 
the correction $\delta D_2$ we obtain a convergent contribution for 
$a> a_{c2}$ 
and a diverging $\delta D_2$ for $a < a_{c2}$.

For $\epsilon >0$ or $a < a_{c1}$~ $\Delta^* \neq 0$ and we obtain
Eqs. (\ref{hom5}) and (\ref{hom6})
diverging in the low frequency and long wavelength limits, respectively.

Finally, we obtain for the temporal and spatial behavior of $D_2$
for $\epsilon <0$
\begin{eqnarray}
D_2 (\bbox{0}, t, D_2, \Delta) 
&\propto& t^{- a/a_{c2}} 
\\
D_2 (\bbox{r}, 0, D_2, \Delta) 
&\propto& r^{a_{c2} - a - d}
\end{eqnarray}
and for $\epsilon > 0$ Eqs. (\ref{hom7}) and (\ref{hom8}).

The scaling properties in the isotropic long range case as compared to the 
short range case are conveniently summarized in  Fig. 13 and Fig. 14.
In  Fig. 13 we have plotted the fall-off exponent as a function of $\mu$.
The lines $a = a_{c2}$ and $a = a_{c1}$ delimit three regions I, II, 
and III.  In I the diffusion coefficient $D_2$ vanishes and the disorder is 
{\it irrelevant}, in II the diffusion coefficient $D_2$ diverges and 
the disorder 
is {\it irrelevant}, and in III $D_2$ diverges and disorder becomes 
{\it relevant}.
We note that Fig. 13 is identical to Fig. 11 with the dimension 
$d$ replaced by the exponent $a$.
In Fig. 14 we have in a plot of the exponent $a$ versus the dimension 
$d$ contrasted the long range case to the short range case.  The short and 
long range cases are delimited by the line $d = a$.  For $a > d$ we have 
the short range case, for $a < d$ the long range case (see Sec. III).  In 
region I delimited by $a = a_{c1}$ and $d = d_{c1}$ $D_2$ diverges and 
the disorder is {\it relevant}, in region II delimited by $a = a_{c2}$ and 
$d = d_{c2}$ $D_2$ diverges and the disorder is {\it irrelevant}, and in 
region I $D_2$ vanishes and disorder is {\it irrelevant}.  In the Brownian 
case $\mu =2$ and region II vanishes; the divergence of $D_2$ coincides with 
the disorder becoming relevant for $d =2$ in the short range case and $a=2$ 
in the long range case, see (\cite{bouch88}).
In Appendix B we briefly summarize the relevant aspects of
the renormalization group flow in the 
$\Delta - D_2$ plane. 

In the case of a general anisotropic force field the 
fixed point structure becomes more complicated and we encounter
the same features as in the Brownian case, refs. \cite{bouch90}
and \cite{bouch88}. The analysis nevertheless proceeds as in the
isotropic case and we therefore defer the discussion to 
Appendix G (the short range case) and Appendix H (the long range case).
\section{Summary and conclusion}
In the present paper we have discussed L\'{e}vy flights in 
$d$ dimensions in a variety
of quenched force fields with fall-off exponent $a$. 
The main conclusion of the paper is that
the long range characteristics of the random motion, characterized by the
dynamic exponent $z$, is essentially not influenced by the quenched
force field and the exponent $z$ locks onto the step index $\mu$ 
in all cases. In other words, the long range character of the
L\'{e}vy steps enables the walker to escape the inhomogeneous
pinning environment created by the quenched force field. 

This 
behavior is entirely different from the case of ordinary Brownian
motion where the dynamic exponent $z$ is enhanced, corresponding
to subdiffusive behavior for $d<2$ in the case of short range
forces and for $a<2$ in the case of long range forces.

Although the dynamic exponent $z$ is unaffected by the quenched 
environment the phenomenon is still characterized by a critical
dimension $d_{c1}$ in the short range case and a critical fall-off
exponent $a_c$ in the long range case. The critical parameters
delimit the relevance of the quenched force field. 
For $d>d_{c1}=2\mu -2$ in the short range case and $a>d_{c1}=2\mu -2$
in the long range case, the force correlations scale to zero,
indicating that the quenched background does not influence the
long range walk characteristics. We note that in the Brownian
case $\mu =2$ and, consequently, $d_{c1}=2$ in the short range case,
coinciding with the dimension where Brownian walk becomes
transparent, i.e., with a finite probability of revisiting a
site and thus becoming sensitive to the force correlations.

For $d<d_{c1}$ or $a<a_c$ the strength of the force correlations
characterized by $\Delta$ scales to a finite (fixed point) value,
showing the relevance of the quenched background. In the
Brownian case this gives rise to a change of the dynamic
exponent $z$, unlike the L\'{e}vy case where $z$ remains
unchanged

A further aspect of the L\'{e}vy case is the appearance of a 
second critical parameter: $d_{c1} =\mu$ in the short
range case and $a_c =\mu$ in the long range case.
These parameters delimit the behavior of the subleading
diffusive term characterized by the ordinary diffusion
coefficient $D_2$, which, of course, in the Brownian case
is the leading term. For $d>d_{c1}$ or $a>a_c$ the wave number
dependent diffusion coefficient $D_2(\bbox{k})$ vanishes
in the long wavelength limit $\bbox{k}\rightarrow 0$;
for $d<d_{c1}$ or $a<a_c$, correspondingly, $D_2(\bbox{k})$
diverges for $\bbox{k}\rightarrow 0$. In the L\'{e}vy
case this behavior is subleading and does not affect
the leading term characterized by the (unchanged) dynamic
exponent $z$. The above behavior is illustrated in
Figs. 11,13, and 14.

In the anisotropic short range and long range cases we
encounter as in the Brownian case a more complicated 
fixed point and renormalization group flow structure,
depicted in Figs 16 and 17. However, as mentioned
above the exponent $z$ remains locked onto $\mu$ and
we find that the behavior of the subleading diffusion
coefficient $D_2$ is the same as in the isotropic
case.

Below in Table 1 we have summarized the behavior of $D_2$
in the various cases.
\begin{table}
\begin{tabular}{ccccc}
&$D_2(\bbox{k})$&$D_2(\bbox{r})$&$D_2(\omega)$&$D_2(t)$\\
\tableline
$d>2\mu-2$&$k^{d-\mu}$&$r^{\mu-2d}$&$\omega^{(d-\mu)/\mu}$&$t^{-d/\mu}$\\
\hline
$d<2\mu-2$&$k^{\mu-2}$&$r^{2-d-\mu}
$&$\omega^{(\mu-2)/\mu}$&$t^{-(2\mu-2)/\mu}$\\
\hline
$a>2\mu-2$&$k^{a-\mu}$&$r^{\mu-d-a}$&$\omega^{(a-\mu)/\mu}$&$t^{-a/\mu}$\\
\hline
$a<2\mu-2$&$k^{\mu-2}$&$r^{2-d-\mu}
$&$\omega^{(\mu-2)/\mu}$&$t^{-(2\mu-2)/\mu}$\\
\end{tabular}
\smallskip
\smallskip
\smallskip
\caption{Table of the behavior of the wave number, space, frequency and
time dependent subleading ordinary diffusion coefficient in the
case of short range and long range force correlations in the
quenched background.
}
\end{table}
\acknowledgements
{
The author wishes to thank K.B. Lauritsen, L. Mikheev, M.H. Jensen,
H.J. Jensen, A. Svane and T. Bohr for useful conversations.
}

\draft
\section{Appendix}
In the ensuing appendices we include more technical aspects of 
our calculations.  
In Appendix A we discuss the derivation of a Fokker-Planck equation for 
L\'{e}vy flights, in Appendix B we derive self energy, vertex, and force 
corrections to first loop order, and in Appendix C we derive renormalization 
group equations in the case of L\'{e}vy flights in a quenched force field.
In Appendix D we consider the renormalization group flow in the isotropic
short and long range cases. Appendix E we show that the dynamical exponent
does not depend on whether we keep the L\'{e}vy coefficient or
diffusion coefficient fixed under scaling. In Appendix F we summarize the
isotropic long range case for Brownian walks. Finally Appendix G and H are 
devoted to a discussion of the anisotropic short and long range cases,
respectively.
\subsection{The Fokker-Planck equation for L\'{e}vy flights}
The usual derivation of the Fokker-Planck equation \cite{fogedby80,strato63} 
depends in an essential way on the existence of 
the moments of the distribution.  
Precisely this assumption breaks down for L\'{e}vy flights for a step size 
distribution with $f < 2$ and we must reconsider the derivation of the 
associated Fokker-Planck equation.  In Sec. II we gave a heuristic derivation 
of the Fokker-Planck equation based on the motion in the absence of a force 
field.  Here we derive the Fokker-Planck equation generalizing the standard 
procedure and relaxing the assumption of finite moments.  Adapting the 
discussion in ref. \cite{strato63} we define the functional

\begin{equation}
I = \int R(\bbox{r}_2) 
\frac{\partial P( \bbox{r}_2 |\bbox{r}_1, t)}{\partial t} d^d r_2~,
\end{equation}
where $P(\bbox{r}_2 | \bbox{r}_1, t)$ is the conditional probability 
distribution and $R(\bbox{r})$ is a generator function.  Using the 
``chain rule'' for a Markovian stochastic process,

\begin{equation}
P(\bbox{r}_2 |\bbox{r}_1, t) = 
\int P(\bbox{r}_2 |\bbox{r}, t_1) P (\bbox{r} |\bbox{r}_1, t-t_1) d^dr 
\end{equation}
and expanding in Fourier modes,
$
R(\bbox{r}) = \sum_{\bbox{k}} e^{i \bbox{kr}} R_{\bbox{k}},
$
we obtain, replacing 
$\partial P/ \partial t$ by $(P(t + \Delta t) - P(t))/\Delta t$, 
\begin{eqnarray}
I=\sum_{\bbox{k}}\int d^d r_2 e^{i\bbox{k\cdot r_2}} 
R_{\bbox{k}}\frac{\partial P(\bbox{r}_2|\bbox{r}_1,t)}{\partial t}=
\sum_{\bbox{k}}\int d^d r e^{i\bbox{k\cdot r}} 
R_{\bbox{k}}
\left[\int d^d r_2 \frac{e^{i \bbox{k\cdot (r_2-r)}}-1}{\Delta t} 
P(\bbox{r}_2|\bbox{r},\Delta t)\right]P(\bbox{r}|\bbox{r}_1,t)~.
\label{a12}
\end{eqnarray}
In order to evaluate
\begin{equation}
\int d^d r_2\frac{e^{i\bbox{k\cdot(r_2-r)}}-1}{\Delta t} 
P (\bbox{r}_2|\bbox{r},\Delta t) = 
\langle\frac{e^{i \bbox{k}\cdot(\bbox{r}(\Delta t)-\bbox{r}(0))} - 1}
{\Delta t}\rangle
\label{a13}
\end{equation}
we invoke the Langevin equation for L\'{e}vy flights (\ref{le}),
$
d\bbox{r}/dt = \bbox{F}(\bbox{r}) + \bbox{\eta}
$
with incremental solution
\begin{equation}
\bbox{r}(\Delta t)-\bbox{r}(0) = {\bbox{F}} (\bbox{r})\Delta t+ 
\bbox{\eta}\Delta t ~.
\label{a14}
\end{equation}
Inserting Eq. (\ref{a14}) in Eq.(\ref{a13}), expanding and using
the result
$
\langle e^{i \bbox{k\cdot\eta}\Delta t} - 1\rangle = 
-\text{const.} k^\mu \Delta t
$
for a L\'{e}vy distribution (cf. Eqs. (\ref{pd}-\ref{plt})) we obtain
$
\left\langle
(e^{i\bbox{k}\cdot 
(\bbox{F}(\bbox{r})\Delta t+\bbox{\eta}\Delta t}-1)/
\Delta t\right\rangle\simeq-\text{const.}k^\mu+i\bbox{k}
\cdot\bbox{F} (\bbox{r})
$
which, inserted in 
in Eq. (\ref{a12}) and requiring it to hold
for any variation of $R$, yields the Fokker-Planck equation 
for L\'{e}vy flights (in momentum space),
\begin{equation}
\frac{\partial P(\bbox{k}, t)}{\partial t} = 
-\text{const.} k^\mu P(\bbox{k}, t)-i\bbox{k}\cdot 
\sum_{\bbox{p}}\bbox{F}(\bbox{k-p}) P (\bbox{p}, t) 
\end{equation}
\subsection{Self energy, vertex, and force corrections to first loop order}
Here we evaluate the self energy, vertex and force corrections to lowest 
order for L\'{e}vy flights in a general anisotropic force field using the 
diagrammatic rules and diagrams in Sec. III.

The self energy correction is given by the diagrams in Fig. 7.  
$\Sigma(\bbox{k},\omega)$ splits up in longitudinal and transverse 
contributions, i.e.,
$
\Sigma(\bbox{k},\omega) = \Sigma_L(\bbox{k},\omega) + 
\Sigma_T(\bbox{k},\omega)~.
$
For $\Sigma_L$ we have
\begin{equation}
\Sigma_L(\bbox{k},\omega)= - 
\lambda_L^2\int\frac{d^d p}{(2\pi)^d}~ 
\Delta^L(\bbox{k}/2 - \bbox{p}) 
\frac{\bbox{k}\cdot 
(\bbox{k}/2-\bbox{p})~
(\bbox{k}/2-\bbox{p})\cdot 
(\bbox{k}/2+\bbox{p})}
{(\bbox{k}/2-\bbox{p})^2}~ 
G_0(\bbox{k}/2+\bbox{p},\omega)~.
\label{a21}
\end{equation}
In order to extract the leading contribution to $O(k^2)$ we first symmetrize 
Eq. (\ref{a21}) by replacing $\bbox{p}$ by $-\bbox{p}$ 
and, furthermore, expanding 
$G_0^{-1}(\bbox{k}/2\pm\bbox{p},\omega)$ 
and $\Delta^L(\bbox{k}/2\pm\bbox{p})$ for small $k$,
$
G_0^{-1}(\bbox{k}/2\pm\bbox{p},\omega) 
= G_0^{-1} 
(\bbox{p},\omega)\pm(\bbox{p\cdot k}/2) 
(\mu D_1 p^{\mu - 2} + 2 D_2 ) 
$
and
$
\Delta^L(\bbox{k}/2\pm\bbox{p}) 
=\Delta^L(\bbox{p})\pm(\bbox{p\cdot k}/2)(a_L-d)p^{a_L-d-2}
$
we obtain to $O(k^2)$

\begin{eqnarray}
&&\Sigma_L (\bbox{k},\omega) =
\nonumber
\\
&&\frac{1}{2} k^2 \frac{\lambda_L^2}{d}\int\frac{d^d p}{(2\pi)^d} 
\Bigl\{[(d-2)G_0^{-1}(\bbox{p},\omega)+ 
(\mu D_1 p^\mu + 2 D_2 p^2)]\Delta^L (\bbox{p}) 
+ G_0^{-1}(\bbox{p},\omega)\Delta_2^L(a_L-d)p^{a_L-d}\Bigr\} 
G_0 (\bbox{p}, \omega)^2  ~.
\end{eqnarray}
Similarly, we find for $\Sigma_T (\bbox{k}, \omega)$ to $O(k^2)$, 

\begin{equation}
\Sigma_T(\bbox{k}, \omega) = 
- \frac{1}{2} k^2 \frac{\lambda^2_T}{d} \int 
\frac{d^dp}{(2\pi)^d} (2d - 2) \Delta^T (\bbox{p}) 
G_0 (\bbox{p}, \omega)~.
\end{equation}
In the static limit $\omega \rightarrow 0$ and in the isotropic short range 
case, $\lambda_L = \lambda_T = \lambda$, and 
$\Delta^{L,T} (\bbox{p}) = \Delta$, we obtain in particular the expression 
yielding Eq. (\ref{dc}),
\begin{equation}
\Sigma(\bbox{k},0) =
-\frac{1}{2} k^2 \frac{\lambda^2}{d} \Delta \int
\frac{d^dp}{(2 \pi)^d}
\frac{D_1 (d-\mu) p^\mu + D_2 (d-2 )p^2}{[D_1 p^\mu + D_2 p^2]^2}~.
\label{a22}
\end{equation}
In the context of the momentum shell integration method discussed in 
Sec. III the dilution of degrees of freedom requires that the internal 
momenta $\bbox{k}/2 + \bbox{p}$ and $\bbox{k} /2 - \bbox{p}$ of the propagator 
and force contraction, respectively, must lie in the momentum shells 
$e^{-\ell} <|\bbox{k}/2 + \bbox{p}|<1$ and 
$e^{-\ell} < | \bbox{k}/2 - \bbox{p} |<1$. We notice that this assignment 
is invariant under the symmetrization $\bbox{p} \rightarrow - \bbox{p}$ 
performed in order to extract the leading $k^2$ correction to $\Sigma$.  On 
the shell we thus obtain for $\Sigma (\bbox{k}, 0)$ in Eq. (\ref{a22}) for 
small $\ell$,

\begin{equation}
\Sigma(\bbox{k}, 0) = 
- \frac{1}{2} k^2 \lambda^2 \Delta \frac{S_d}{(2 \pi)^d} \frac{1}{d} 
\frac{D_1 (d-\mu) + D_2 (d-2)}{[D_1 + D_2]^2}\ell,
\end{equation}
where 
$S_d = 2\pi^{d/2}/(d/2-1)!$ 
(cf. Eq. (\ref{ld})), leading to Eq. (\ref{dcor}) for the correction 
$\delta D_2$ to the diffusion coefficient.

In a completely similar manner we obtain in the general case on the shell 
$p = 1$,

\begin{eqnarray}
& &\Sigma(\bbox{k}, 0)= \nonumber\\
& & \frac{1}{2}k^2\frac{S_d}{(2\pi)^d}\frac{1}{d}
\frac{\lambda_L^2 
\left[(d+\mu)D_1(\Delta_1^L+\Delta_2^L)+(a_L-d)(D_1 + D_2) 
\Delta_2^L\right]-\lambda_T^2(2d-2)(D_1+D_2)
(\Delta_1^T + \Delta_2^T)}{[D_1 + D_2]^2} \ell ~.
\end{eqnarray}
The vertex corrections are given by the diagrams in Fig. 8.  In the 
isotropic short range case the correction is given in Eq. (\ref{vc}) and 
$\delta \lambda$ in Eq. (\ref{vc1}) follows directly, i.e., no 
symmetrization is required here since the correction in Eq. (\ref{vc}) is 
already of order $k$.  Focusing on the corrections to $\lambda_L$ and 
$\lambda_T$ the calculation is also quite simple in the general case.  
We obtain

\begin{eqnarray}
\delta \lambda_T =&&
- \lambda_T \lambda_L^2 \frac{1}{d} \int 
\frac{d^dp}{(2 \pi)^d}~ \Delta^L (\bbox{p}) G_0 (\bbox{p}, 0)^2
\\
\delta \lambda_L =&&
- \lambda_L^3 \frac{1}{d} \int
\frac{d^d p}{(2 \pi)^d} \Delta^L (\bbox{p}) G_0 (\bbox{p}, 0)
\end{eqnarray}
and on the shell $e^{-\ell} < p < 1$
\begin{eqnarray}
\delta \lambda_T =&& 
- \lambda_T \lambda_L^2 \frac{S_d}{(2 \pi)^d} 
\frac{1}{d} \frac{\Delta_1^L + \Delta_2^L}{[D_1 + D_2]^2} \ell
\\
\delta \lambda_L =&&
- \lambda_L^3 \frac{S_d}{(2 \pi)^d}
\frac{1}{d} \frac{\Delta_1^L + \Delta_2^L}{[D_1 + D_2]^2} \ell ~ .
\end{eqnarray}

The force corrections are given by the diagrams in Fig. 9.  In the isotropic 
short range case the results are given in Eqs. (\ref{vc2}-\ref{nc}).  In 
the general case we obtain the correction to the vertex function 
$\Gamma (\bbox{k},\bbox{p},\bbox{l} ,\omega)$ defined diagrammatically in 
Fig. 6, introducing the tensors (dyadics)
$\tilde{\Delta}^L (\bbox{k}) =
[k^{\alpha} k^{\beta}/ k^2]\Delta^L(\bbox{k})$ and
$\tilde{\Delta}^T(\bbox{k}) =
[\delta^{\alpha \beta} - k^{\alpha} k^{\beta} /k^2]\Delta^T(\bbox{k})$,
\begin{eqnarray}
\Gamma (\bbox{k},\bbox{p},\bbox{l}, \omega )
= 
&-&\lambda_L^2\bbox{k}\cdot\tilde{\Delta}^L(\bbox{l})\cdot\bbox{p} - 
\lambda_T^2\bbox{k}\cdot\tilde{\Delta}^T(\bbox{l})\cdot\bbox{p}\nonumber\\
&+& 
\lambda_L^4\int\frac{d^d n}{(2\pi)^d}\bbox{k}\cdot 
\tilde{\Delta}^L(\bbox{n})\cdot
(\bbox{p}+\bbox{l}-\bbox{n})
(\bbox{k}-\bbox{n})\cdot\tilde{\Delta}^L(\bbox{l}-\bbox{n})\cdot 
\bbox{p}
G_0(\bbox{k}-\bbox{n},\omega) 
G_0(\bbox{p}+\bbox{l}-\bbox{n},\omega)\nonumber\\
&+& 
\lambda_T^4 \int\frac{d^dn}{(2 \pi)^2}\bbox{k}\cdot 
\tilde{\Delta}^T(\bbox{n})\cdot(\bbox{p}+ \bbox{l} - \bbox{n})
(\bbox{k} - \bbox{n})\cdot\tilde{\Delta}^T(\bbox{l}-\bbox{n})\cdot 
\bbox{p}
G_0 (\bbox{k}-\bbox{n},\omega) 
G_0(\bbox{p}+\bbox{l}-\bbox{n},\omega)\nonumber\\
&+& 
\lambda_L^2 \lambda_T^2\int\frac{d^dn}{(2\pi)^d}\bbox{k}\cdot 
\tilde{\Delta}^L (\bbox{n})\cdot(\bbox{p}+\bbox{l}-\bbox{n})
(\bbox{k}-\bbox{n})\cdot\tilde{\Delta}^T (\bbox{l}-\bbox{n})\cdot\bbox{p} 
G_0(\bbox{k}-\bbox{n},\omega)
G_0(\bbox{p}+\bbox{l}-\bbox{n},\omega)\nonumber\\
&+&
\lambda_T^2\lambda_L^2\int\frac{d^d n}{(2\pi)^d}\bbox{k}\cdot
\tilde{\Delta}^T(\bbox{n})\cdot(\bbox{p}+\bbox{l}- \bbox{n})
(\bbox{k}-\bbox{n})\cdot\tilde{\Delta}^L(\bbox{l}-\bbox{n})\cdot 
\bbox{p} 
G_0(\bbox{k}-\bbox{n},\omega) 
G_0(\bbox{p}+\bbox{l}-\bbox{n},\omega)\nonumber\\
&+& 
\lambda_L^4\int\frac{d^d n}{(2 \pi)^d}\bbox{k}\cdot 
\tilde{\Delta}^L (\bbox{n})\cdot\bbox{p}
(\bbox{k}-\bbox{n})\cdot\tilde{\Delta}^L(\bbox{l}-\bbox{n})\cdot 
(\bbox{p}+\bbox{n}) 
G_0(\bbox{k}-\bbox{n},\omega)
G_0(\bbox{p}+\bbox{n},\omega)\nonumber\\
&+& 
\lambda_T^4
\int\frac{d^d n}{(2 \pi)^d} 
\bbox{k}\cdot\tilde{\Delta}^T (\bbox{n})\cdot\bbox{p}
(\bbox{k}-\bbox{n})\cdot 
\tilde{\Delta}^T(\bbox{l}-\bbox{n})\cdot(\bbox{p}+\bbox{n}) 
G_0(\bbox{k}-\bbox{n},\omega)
G_0(\bbox{p}+\bbox{n},\omega)\nonumber\\
&+& 
\lambda_L^2\lambda_T^2\int\frac{d^dn}{(2\pi)^d} 
\bbox{k}\cdot\tilde{\Delta}^L (\bbox{n})\cdot\bbox{p}
(\bbox{k}-\bbox{n})\cdot\tilde{\Delta}^T (\bbox{l}-\bbox{n})\cdot 
(\bbox{p}+\bbox{n})
G_0 (\bbox{k}-\bbox{n},\omega) 
G_0(\bbox{p}+\bbox{n},\omega)\nonumber\\
&+&
\lambda_T^2\lambda_L^2\int\frac{d^dn}{(2\pi)^d}
\bbox{k}\cdot\tilde{\Delta}^T (\bbox{n})\cdot\bbox{p}
(\bbox{k}-\bbox{n})\cdot\tilde{\Delta}^L (\bbox{l}-\bbox{n})\cdot
(\bbox{p}+\bbox{n})
G_0 (\bbox{k}-\bbox{n},\omega)
G_0(\bbox{p}+\bbox{n},\omega) ~.
\label{lver}
\end{eqnarray}

In order to identify for example the correction to $\Delta_1^T$ we choose 
$\bbox{k}\cdot\bbox{l} = 0$ and take the $\bbox{l}\rightarrow 0$ limit.  
Using that $\bbox{k}\cdot\tilde{\Delta}^L 
(\bbox{l})\cdot\bbox{p}=0, \bbox{k}\cdot\tilde{\Delta}^T 
(\bbox{l})\cdot\bbox{p}=\bbox{k}\cdot\bbox{p}\Delta^T 
(\bbox{l}),\bbox{k}\cdot\tilde{\Delta}^T(\bbox{n})\cdot\bbox{n} 
= 0,$ and 
$\bbox{k}\cdot\tilde{\Delta}^L (\bbox{n})\cdot\bbox{n}= \bbox{k}\cdot 
\bbox{n} \Delta^T (\bbox{n})$, the expression (\ref{lver}) reduces 
considerably and we find the correction

\begin{equation}
\delta \Delta_1^T = 
\lambda_L^2 \left(1-\frac{1}{d}\right) 
\int \frac{d^dp}{(2 \pi)^d} \Delta^T (\bbox{p}) 
\Delta^L (\bbox{p}) G_0^2 (\bbox{p}, 0)
\end{equation}
and on the shell $e^{-\ell} < p< 1$, 
\begin{equation}
\delta \Delta_1^T = \lambda_L^2 \frac{S_d}{(2 \pi)^d} 
\left(1 - \frac{1}{d}\right) \frac{(\Delta_1^T + \Delta_2^T )
(\Delta_1^L + \Delta_2^L)}{[D_1 + D_2]^2} \ell~.
\end{equation}
Similarly, we find the correction to $\Delta^L$ by 
choosing $\bbox{k} = \bbox{l}$, 

\begin{equation}
\delta \Delta_1^L = \delta \Delta_1^T~,
\end{equation}
i.e., the force corrections in the longitudinal and transverse 
cases are identical.
\subsection{Renormalization group equations in the case 
of a general force field}

Here we derive the renormalization group equations for a general 
force field to first loop order following the procedure in Sec. III.
Including the corrections to $D_2, \lambda_L, \lambda_T, \Delta^L$ and 
$\Delta^T$ we obtain the renormalized Fokker-Planck equation and force 
correlations (cf. Eqs. (\ref{eq1}-\ref{eq2})),

\begin{eqnarray}
(-i\omega+D_1k^\mu+(D_2+\delta D_2)k^2) 
P(\bbox{k},\omega) 
=&&P_0(\bbox{k})
+(\lambda_L+\delta\lambda_L)\int\frac{d^dp}{(2\pi)^d}(-i)\bbox{k}\cdot 
{\bbox{E}}(\bbox{k}-\bbox{p})P(\bbox{p},\omega)\nonumber
\\
+&&(\lambda_T+\delta\lambda_T)\int 
\frac{d^dp}{(2\pi)^d}(-i)\bbox{k}\cdot 
{\bbox{B}}(\bbox{k}-\bbox{p})P(\bbox{p},\omega)
\\
\langle E^\alpha (\bbox{k}) E^\beta (\bbox{p}) \rangle_F 
=&& \left(\Delta^L (\bbox{k}) + \delta \Delta^L 
(\bbox{k})\right)\left[ \frac{k^\alpha k^\beta}{k^2}\right] 
(2 \pi)^d \delta (\bbox{k} + \bbox{p})
\\
\langle B^\alpha (\bbox{k}) B^\beta (\bbox{p}) \rangle_F 
=&& \left(\Delta^T (\bbox{k}) + \delta \Delta^T 
(\bbox{k})\right) \left[ \delta^{\alpha \beta} - 
\frac{k^\alpha k^\beta}{k^2} \right] (2 \pi)^d \delta 
(\bbox{k} + \bbox{p}) ~,
\end{eqnarray}
where $\bbox{k}$ and $\bbox{p}$ after the momentum shell integration now 
range in the interval $1 < k,p < e^{-\ell}$.  
Renormalizing wave numbers, frequency, and distribution as in 
Eqs(\ref{rs1}-\ref{rs3}) and $\bbox{E}$ and $\bbox{B}$ according to
${\bbox{E}}^\prime(\bbox{k}^\prime)= 
\exp(-\beta_L(\ell)){\bbox{E}}(\bbox{k})$, 
and ${\bbox{B}} (\bbox{k}^\prime)=\exp(-\beta_T(\ell)){\bbox{B}}(\bbox{k})$, 
where $\alpha$ and $\beta_{L,T}$ are given 
by Eqs. (\ref{alfa}-\ref{beta}) with $u$ replaced by $u_{L,T}$
we obtain a renormalized Fokker-Planck equation and force correlations 
of the same form as in Eqs. (\ref{rfp}-\ref{rfc}).
Furthermore, from  the expressions in Eqs. (\ref{fcf1}-\ref{fcf2}) for 
$\Delta^T (\bbox{k})$ and $\Delta^L (\bbox{k})$ 
we identify scale dependent parameters and obtain
the renormalization group equations
\begin{eqnarray}
D_1^\prime(\ell)
&=& 
D_1e^{-\mu\ell+\alpha}
\\
D_2^\prime(\ell)
&=& 
(D_2+\delta D_2)e^{-2\ell}
\\
\lambda_L^\prime (\ell)
&=& 
(\lambda_L+\delta\lambda_L)e^{-(1+d)\ell+\alpha+\beta_L}
\\
\lambda_T^\prime(\ell)
&=&
(\lambda_T+\delta\lambda_T)e^{-(1+d)\ell+\alpha+\beta_T}
\\
\Delta_1^{L^\prime}(\ell)
&=&
(\Delta_1^L+\delta\Delta_1^L)e^{d\ell- 2\beta_L}
\\
\Delta_2^{L^\prime}(\ell)
&=& 
\Delta_2^Le^{d\ell-2\beta_L-\ell(a_L-d)}
\\
\Delta_1^{T^\prime}(\ell)
&=&
(\Delta_1^T+\delta\Delta_1^T)e^{d\ell-2\beta_T}
\\
\Delta_2^{T^\prime}(\ell)
&=&
\Delta_2^Te^{d\ell-2\beta_T-\ell(a_T-d)}
\end{eqnarray}
or in differential form the Eqs. (\ref{rggg1}-\ref{rggg6}).
We have here again fixed the vertex constants to $\lambda_{L,T}=1$
by choosing 
$
u_L 
= 
1+d-z-\lambda_L^{-1}\delta\lambda_L/{\ell}
$
and 
$
u_T 
=
1+d-z-\lambda_T^{-1}\delta\lambda_T/{\ell}
$.
\subsection
{
The renormalization group flow in the
isotropic case
}
\subsubsection
{
The 
isotropic short range case
}

Eliminating 
$\ell$ in Eqs. (\ref{sol1})
and (\ref{sol2}) 
and setting $\mu=d_{c_2}$ and $\epsilon=d_{c_1}-d$ we obtain 
\begin{equation}
\delta D_2 =
\left[\delta D^0_2 + A \frac{\delta \Delta^0}{D_1} \right]
\left[ \frac{\delta \Delta}{\delta \Delta^0}\right]
^{(2 - d_{c_2})/|d_{c_1}-d|} -
\frac{A}{D_1} \delta \Delta~.
\label{m1}
\end{equation}
In Fig. 15 we have depicted the renormalization group flow about the
Gaussian fixed point $(\Delta^*, D_2^*) = (0,0)$ in the three cases: 
a) $d>d_{c2}$, b) $d= d_{c2}$, and c) $d_{c2} > d> d_{c1}$.
In the case a) for
$d>d_{c_2}$, corresponding to region I in Fig. 11, the first term
in Eq. ({\ref{m1}) dominates and near the fixed point
$
\delta D_2 \sim
[\delta D_2^0 + A\delta\Delta^0/D_1]
[\delta\Delta/\delta\Delta^0]^{(2 - d_{c_2})/|d_{c_1}-d|}
$,
i.e., the trajectories approach the fixed point with vertical slope except for
the trajectories originating from the domain of initial values on the line
$\delta D^0_2 = - (A/D_1)\delta \Delta^0$, corresponding to the position of
the non-trivial fixed point emerging below $d_{c1}$.  In this regime $D_2$
vanishes as discussed above.
In the case b) for
$d=d_{c_2}$, the marginal case, we have
$\delta D_2 = (\delta D_2^0/\delta\Delta^0)\delta\Delta$
and the trajectories approach the fixed point with constant slope, i.e.,
the linear scaling regime depends on $\mu$ and becomes largest for $\mu = d$,
precisely the case where naive perturbation theory or the scaling analysis
above yield a divergent $D_2$.
In the case c) for
$d_{c1} < d< d_{c2}$, corresponding to region II in Fig. 11,
we have $\delta D_2=-(A/D_1)\delta\Delta$.
The trajectories now approach the fixed point with a constant slope
$- A/D_1$, except for trajectories with $\delta \Delta^0 = 0$ which
lie on the line $\Delta = 0~.$
\subsubsection
{
The 
isotropic long range case
}
Eliminating $\ell$ in the renormalization group
equations (\ref{soll1}-\ref{soll2}) we have
 
\begin{eqnarray}
\delta D_2 = \left[ \delta D_2^0 +
A \frac{3 \mu - 2 d -2}{2-\mu} \frac{\delta \Delta^0}{D_1} \right]
\left[ \frac{\delta \Delta}
{\delta \Delta^0}\right]^{( 2-d_{c2})/|a_{c_1}-a|}
 - A \frac{3 \mu - 2d - 2}{2-\mu} \delta \Delta \ .
\end{eqnarray}
Focusing on the flow about the Gaussian fixed point
$(\Delta^*, D_2^*) = (0,0)$ the discussion in the isotropic
short range case above applies with $d$
replaced by $a$.  For  $d > a> \mu$ the
trajectories approach the fixed point with vertical slope except for the
trajectories originating from the domain of initial values on the line
$\delta D_2^0=-\left(A(3\mu-2d-2)\left/(2-\mu)D_1\right)\right.\delta\Delta^0$,
corresponding to the position of the non-trivial fixed point emerging below
$2\mu - 2$.  In the marginal case  $\mu = a$ the
trajectories approach the fixed point with constant slope, i.e., the linear
scaling regime depends on $\mu$ and becomes largest for $\mu = a$, precisely
the case where naive perturbation theory or the scaling analysis above yield
a divergent $D_2$.  For $a < \mu$ the trajectories
approach the fixed point with constant slope $-A (3 \mu - 2d - 2)/(2- \mu)D_1$,
except for trajectories with $\delta \Delta^0 = 0$ which lie on the line
$\Delta = 0$.  Fig. 15 also applies in the long range case with $d$
replaced by $a$ but note that the slope
$- A/D_1$ is replaced by $- A (3\mu - 2d-2) / (2 - \mu) D_1$.
\subsection
{Keeping the ordinary diffusion coefficient constant
under renormalization
}

We here briefly repeat our renormalization group analysis
but now keeping the diffusion coefficient $D_2$ fixed under scaling.
Limiting our discussion to the behavior near the Gaussian fixed point
$\Delta^* = 0$ we thus obtain from Eqs. (\ref{rgg1}-\ref{rgg3})
$
z = 2
$
and the renormalization group equations
$
dD_1/d\ell =
(2 - \mu)D_1
$
and 
$
d\Delta/d\ell =(2 - d)\Delta
$
with solutions $D_1 = D_1^0 \exp((2 - \mu)\ell)$ and
$\Delta = \Delta^0 \exp((2 - d)\ell)$.  From the homogeneity relation
(\ref{hom1}) we have, setting $\alpha(\ell) = 2\ell$,
$
P(\bbox{k}, \omega, D_1, \Delta) =
e^{2\ell}P(\bbox{k}e^{\ell},\omega
e^{2\ell},D^0_1e^{(2-\mu)\ell},\Delta^0 e^{(2 - d)\ell})
$,
which forms the basis for our scaling analysis.  Choosing
$\bbox{k} e^{\ell} \simeq 1$ we deduce the scaling form
$
P(\bbox{k}, \omega, D_1, \Delta) =
k^{-2} P (1, \omega/ k^2, D^0_1k^{\mu - 2}, \Delta^0 k^{d-2})
$
.
Note, however, that this form does not imply $z = 2$ since
$D_1^0k^{\mu - 2}$ and $\Delta^0 k^{d - 2}$ diverge in the long wavelength
limit $k\rightarrow 0$.  In fact choosing $k$ such that
$\Delta^0 k^{d -2} \ll 1$ we obtain perturbatively,
$
P(\bbox{k},\omega,D_1, \Delta) =
k^{-2}(-i\omega/k^2+D_1^0k^{\mu-2}+D_2+\text{const.}\Delta^0 k^{d-2})^{-1}
$
or
$
P(\bbox{k},\omega,D_1, \Delta) =
(-i \omega + D_1^0 k^{\mu} + D_2 k^2 + \text{const.}\Delta^0 k^d)^{-1}
$
similar to our previous results and implying the dynamical exponent $z=\mu$.
\subsection{The isotropic long range case for Brownian walks}

Here we recover the well-known results in the Brownian
case to first loop order \cite{bouch90,bouch88}.  The renormalization group
equations in the isotropic long range Brownian case are extracted from the
general equations (\ref{rggg1}-\ref{rggg6})
by
setting
$\mu=2,D_1=0,D_2 = D,\Delta_1^{L,T}=0,~\Delta_2^{L,T}=\Delta$ , and
$a_{L,T} = a$, i.e.,
\begin{eqnarray}
\frac{dD}{d\ell}&=& (z-2) D +
A (2d-a-2) \frac{\Delta}{D}
\\
\frac{d \Delta}{d\ell} &=& (2z - 2-a) \Delta - 4 A \frac{\Delta^2}{D^2}~.
\end{eqnarray}
Keeping $D$ fixed we have
$
z=2-A(2d - a-2)\Delta/D^2
$
and the equation for $\Delta$,
\begin{equation}
\frac{d \Delta}{d\ell} = (2 - a) \Delta - 2 A (2d - a) \frac{\Delta^2}{D^2}~.
\end{equation}
The expansion parameter is $2-a$.  For $2-a <0$ we obtain the Gaussian fixed
point $\Delta^* = 0$, the disorder is irrelevant and $z$ locks onto 2,
characterizing ordinary diffusion.  For $2 - a > 0 $ we have the non-trivial
fixed point $\Delta^* = (2-a) D^2 / 2 A (2d-a)$ and the dynamic exponent
\begin{equation}
z = 2- \frac{d-2}{d-1} (2-a)~,
\label{EZ}
\end{equation}
signalling anomalous diffusion \cite{bouch88}, due to the interference of the
quenched long range force correlations with the Brownian walk.  In the
vicinity of either fixed point we have
$\Delta(\ell)=\text{const.}\exp(-|2-a|\ell)+\Delta^*$ and we obtain scaling
relations for the distribution $P(\bbox{k},\omega,\Delta)$ and the wave number
and frequency dependent diffusion coefficient $D(\bbox{k},\omega, \Delta)$,
\begin{eqnarray}
P(\bbox{k}, \omega, \Delta)&& =
e^{z\ell} P\left(\bbox{k} e^{\ell}, \omega e^{z\ell}, \Delta^* +
\text{const.} e^{-|2-a|\ell}\right)
\label{E1}
\\
D(\bbox{k}, \omega, \Delta)&& =
e^{(2-z)\ell} D\left(\bbox{k} e^{\ell}, \omega e ^{z\ell}, \Delta^* +
\text{const.} e^{-|2-a|\ell}\right)~.
\label{E2}
\end{eqnarray}
Similar to our discussion in Sec. IV we infer from Eq. (\ref{E1}) the
dynamic exponent $z$ and from Eq. (\ref{E2}), choosing
$\omega e^{z\ell} \simeq 1$ and $\bbox{k} e^{\ell} \simeq 1$,
$
D(\bbox{0},\omega, \Delta)
\propto \omega^{-(2-z)/z}~
\left[\Delta^* + \text{const.} \omega^{-|2-a|/z} \right]
$
and
$
D(\bbox{k}, 0, \Delta)
\propto k^{- (2-z)}~\left[\Delta^* + \text{const.} k^{|2-a|} \right]
$
.
For $a >2 $ we have $\Delta^* = 0$ and $z = 2,$, i.e.,
\begin{eqnarray}
D(\bbox{0},\omega, \Delta)
&\propto& \omega^{(a-2)/2}
\\
D(\bbox{k}, 0, \Delta)
&\propto& k^{a-2}
\end{eqnarray}
which vanish in the $\bbox{k} \rightarrow 0$ and $\omega \rightarrow 0$ limits,
and for the temporal and spatial behavior,
\begin{eqnarray}
D(\bbox{0},t,\Delta)
&\propto& t^{-a/2}
\\
D(\bbox{r}, 0, \Delta)
&\propto& r^{2-a-d}~.
\end{eqnarray}
For $a <2$ we have $\Delta^* >0$ and
$z$ given by Eq. (\ref{EZ}) and
\begin{eqnarray}
D(\bbox{0},\omega,\Delta)
&\propto&
\Delta^*\omega^{-\frac{1}{2}\left(\frac{d-2}{d-1}\right)(2-a)}
\\
D(\bbox{k}, 0, \Delta)
&\propto& \Delta^* \bbox{k}^{- \left( \frac{d-2}{d-1}\right) (2-a)}~,
\end{eqnarray}
where the behavior of $D$ depends also on the dimension of the system
$d$ \cite{bouch90,bouch88}.  Also,
\begin{eqnarray}
D(\bbox{0},t,\Delta)
&\propto& \Delta^*t^{\frac{1}{2}\left(\frac{d-2}{d-1}\right)(2-a)-1}
\\
D(\bbox{r},0,\Delta)
&\propto& \Delta^* r^{\left(\frac{d-2}{d-1}\right) (2-a)-d}~.
\end{eqnarray}
%
\subsection{The anisotropic short range case}
Generally the quenched force field must be expected to be anisotropic 
consisting of a transverse divergence-free part $\bbox{B}$ and a longitudinal 
curl-free part $\bbox{E}$.  The vector character of ${\bbox{F}}$ is 
reflected in 
the transverse 
and
longitudinal correlation functions $\Delta^T$ and 
$\Delta^L$, respectively,
introduced in Sec. II.  The case of Brownian 
motion in a short range anisotropic force field has been discussed to first 
loop order in refs. \cite{aronovitz84,fisher85}.  
As in the isotropic case treated 
in refs. \cite{fisher84,luck83} to second loop order, the critical dimension is 
$d_{c1}= 2;$
below $d_{c1}$ the long time behavior is controlled by the isotropic fixed 
point $\Delta_T^* = \Delta_L^* = \Delta^*$, giving rise to anomalous 
subdiffusion.  However, at intermediate times the diffusional character is 
controlled by a transverse fixed point 
$\Delta_T^* \neq 0, \Delta_L^* = 0$.
\subsubsection{The Brownian case}
In order to clearly illustrate how the L\'{e}vy case differs from the 
Brownian case we briefly discuss the renormalization group equations in the 
Brownian case.  Setting $D_1 = 0, D_2 = D, \Delta_1^{L,T} = \Delta_{L,T}$, 
and $\Delta_2^{L,T} = 0$, we obtain from the general equations 
(\ref{rggg1}-\ref{rggg6})
\begin{eqnarray}
\frac{dD}{d\ell}
&=& 
(z - 2) D + A \frac{(2d - 2) \Delta_T - d \Delta_L}{D^2}~. 
\\[.25cm]
\frac{d \Delta_T}{d\ell}
&=&
(2z - d-2) \Delta_T - 2 A (3-d) \frac{\Delta_L \Delta_T}{D^2}~. 
\\ [.25cm]
\frac{d \Delta_L}{d\ell}
&=&
(2z - d-2) \Delta_L - 4 A \frac{\Delta_L^2}{D^2} + 
2A (d-1) \frac{\Delta_L \Delta_T}{D^2}~.
\end{eqnarray}
Keeping $D$ fixed, we choose
$
z = 2 - A(2(d-1)\Delta_T-d\Delta_L)/D^2
$
and we have the equations for $\Delta_T$ and $\Delta_L$, 
\begin{eqnarray}
\frac{d \Delta_T}{d\ell}
&=& 
(2-d) \Delta_L - 2 \Delta \frac{2(d -1) 
\Delta_T^2 + (3-2 d) \Delta_L \Delta_T}{D^2} 
 \\
\frac{d \Delta_L}{d\ell}
&=& 
(2 - d) \Delta_L - 2 \Delta \frac{(d - 1) 
\Delta_L \Delta_T + (2-d) \Delta_L^2}{D^2}~.
\end{eqnarray}

Above the critical dimension $d_{c1} = 2$ we obtain the stable Gaussian fixed 
point $(\Delta_L^*, \Delta_T^*) = (0,0)$, corresponding to normal diffusion 
for $z = 2$ and irrelevance of the quenched force field.  Below $d_{c1}$ to 
$O(2-d)$ we obtain the stable isotropic fixed point
$
(\Delta_L^*,\Delta_T^*)=[(2-d)D^2/2 A,(2-d)D^2/2A],
$
controlling the long time anomalous behavior characterized by 
$z = 2+2(2-d)^2$ evaluated to second loop order $O((2-d)^2)$ 
\cite{fisher84,luck83} and the unstable anisotropic transverse fixed point 
$(\Delta_L^*, \Delta_T^*) = (0,(2-d)D^2/4 A)$, determining the cross-over at 
intermediate times.  We shall not pursue the Brownian case further here but 
refer to 
refs. \cite{aronovitz84,fisher85,bouch88,honk88,honkonen,honk89,honk91}.
\subsubsection{The L\'{e}vy case}

In the L\'{e}vy case we proceed as in Section IV, setting $z = \mu$ 
in order to fix 
$D_1, \Delta_1^{L,T} = \Delta_{L,T}$, and $\Delta_2^{L,T} = 0$, the general 
equations (\ref{rggg1}-(\ref{rggg6}) imply the renormalization group equations

\begin{eqnarray}
\frac{dD_2}{d\ell}
&=&
(\mu - 2) D_2 
+ A \frac{(D_1 (2-\mu - d) - D_2d) \Delta_L + 
(D_1+D_2) (2d - 2) \Delta_T}{(D_1 + D_2)^2} 
\label{h1}
\\ 
\frac{d \Delta_L}{d\ell}
&=&  
\epsilon \Delta_L - 2 A \frac{2 \Delta_L^2 - (d - 1) 
\Delta_L \Delta_T}{[D_1 + D_2]^2} 
\label{h2}
\\ 
\frac{d \Delta_T}{d\ell}
&=&  
\epsilon \Delta_T - 2 A (3-d) \frac{\Delta_L \Delta_T}{[D_1 + D_2]^2}~,
\label{h3}
\end{eqnarray}
where $\epsilon = d_{c1} - d = 2 \mu - 2 - d$.

Again we observe that irrespective of the vector character of the random 
force $\bbox{F}$ the dynamic exponent $z$ locks onto the L\'{e}vy index 
$\mu$, owing to the long range character of the L\'{e}vy steps.  The 
equations (\ref{h1}-\ref{h3}) also have three fixed points:  
\begin{eqnarray}
(D_2^*, \Delta_L^*, \Delta_T^*) =&& (0,0,0)
~~~~~~~~~~~~~~~~~~~~~~~~~~~~~~~~~~~~~~~~~~~~~~\text{(Gaussian)}
\label{p1}
\\
(D_2^*, \Delta_L^*, \Delta_T^*) =&& \left(- \epsilon \frac{D_1}{2(3-d)}, 
\epsilon \frac{D_1^2}{2 A(3-d)}, \epsilon \frac{D_1^2}{2 A (3 - d)} \right)
~~~~\text{(isotropic)}
\label{p2}
\\
(D_2^*, \Delta_L^*, \Delta_T^*) =&& \left( \frac{\epsilon}{4} 
\frac{D_1 (4-3 \mu)}{2 - \mu}, \epsilon \frac{D_1^2}{4A}, 0\right)
~~~~~~~~~~~~~~~~~~~~~~~\text{(longitudinal)}
\label{p3}
\end{eqnarray}
In order to examine the stability of the fixed points we derive 
renormalization group equations to $O(\epsilon)$.  Setting 
$D_2(\ell)=D_2^* +\delta D(\ell),\Delta_L(\ell)=
\Delta_L^*+\delta\Delta_L(\ell)$, and $\Delta_T (\ell) = 
\Delta_T^* + \delta \Delta_T (\ell)$, we have 
\begin{eqnarray}
\frac{d\delta D_2}{d\ell} 
=&&
\left(\mu-2+A\frac{(d+2\mu-4) 
\Delta_L^*-(2d-2)\Delta_T^*}{D_1^2}\right)\delta 
\Delta_2 
\nonumber\\
&&
+A\frac{D_1(2-\mu-d)+D_2^*(2\mu+d-4)}{D_1^2} 
\delta\Delta_L 
\nonumber\\
&&
+A\frac{(2d-2)(D_1-D_2^*)}{D_1^2}\delta
\Delta_T
\label{k1}
\\ 
\frac{d\delta\Delta_L}{d\ell} 
=
&&
\left(\epsilon-2A 
\frac{4\Delta_L^*-(d-1)\Delta_T^*}
{D_1^2} \right)\delta\Delta_L 
+\left(\frac{2A(d-1)\Delta_L^*}{D_1^2} \right) 
\delta \Delta_T 
\label{k2}
\\ 
\frac{d\delta\Delta_T}{d\ell} 
=
&&
\left(\epsilon-2A(3-d) 
\frac{\Delta_L^*}{D_1^2}\right)\delta\Delta_T 
-\left( 2A(3-d)\frac{\Delta_T^*}{D_1^2}\right)\delta\Delta_L~. 
\label{k3}
\end{eqnarray} 

In the vicinity of the Gaussian fixed point we have the solution

\begin{eqnarray}
\delta D_2 (\ell) 
=&&
\left(\delta D_2^0+\frac{A}{D_1} 
\left(\frac{2-\mu-d}{d-\mu} 
\delta \Delta_L^0+\frac{2(d-1)}{d-\mu} 
\delta \Delta_T^0\right)\right)e^{- (2-\mu)\ell}\nonumber\\ 
&&-\frac{A}{D_1}\left(\frac{(2-\mu-d)}{d-\mu} 
\delta\Delta L^0+\frac{2(d - 1)}{d -\mu} 
\delta\Delta_T^0\right)e^{\epsilon\ell}
\\ 
\delta \Delta_L (\ell) 
=&& \delta \Delta_L^0 e^{\epsilon \ell} 
\\ 
\delta \Delta_R (\ell)
=&& \delta \Delta_T^0 e^{\epsilon \ell} 
\end{eqnarray}
and the fixed point is stable for 
$\epsilon < 0$, i.e., $d > d_{c1} = 2 \mu - 2$.

Eliminating $\ell$ we obtain for the flow in the $(D_2, \Delta_L, \Delta_T)$ 
space for $\epsilon < 0$

\begin{eqnarray}
\delta D_2 =&&
\left[\delta D_2^0+\frac{A}{D_1} 
\left(\frac{2-\mu-d}{d-\mu} 
\delta\Delta_L^0+\frac{2(d-1)}{d-\mu}\delta\Delta_T^0 \right) 
\right]
\left[ \frac{\delta \Delta_{L,T}}
{\delta \Delta_{L,T}^0} \right]^
{(2 - \mu/|\epsilon|}\nonumber\\ 
-&&\frac{A}{D_1} \left[ \frac{2 -\mu -d}{d -\mu} 
\delta \Delta_L + \frac{2(d-1)}{d-\mu} \delta \Delta_T \right] 
\end{eqnarray}
and the discussion in the isotropic case applies with a few modifications.  
For $d > d_{c2}$ the trajectories approach the fixed point with vertical slope 
with exception of trajectories lying in the plane

\begin{equation}
\delta D_2 = 
- \frac{A}{D_1} 
\left[ \frac{2 - \mu - d}{d - \mu} \delta \Delta_L + 
\frac{2(d-1)}{d - \mu} \delta \Delta_T \right]~.
\label{s1}
\end{equation}

In the limiting case $d = \mu$ the trajectories approach the fixed point 
with constant slope, i.e., the linear scaling regime depends on $\mu$ and 
becomes largest for $\mu = d$, corresponding to the case where naive 
perturbation theory yields a divergent diffusion coefficient.  Finally, 
for $d_{c1} < \mu < d$ the flow approaches the fixed point tangentially to the 
plane defined by Eq. (\ref{s1}).  The characteristics of the flow are 
depicted in Fig. 15.

The scaling analysis of $P$ and $D_2$ also proceeds as in the isotropic case. 
Suppressing the dependence on $D_1$ which is kept fixed we define
$
P(\bbox{k}, \omega, D_2, \Delta_L, \Delta_T) = 
\langle P (\bbox{k}, \omega) \rangle_F
$
and we obtain the homogeneity relations: 

\begin{eqnarray}
P(\bbox{k}, \omega, D_2, \Delta_L, \Delta_T) =&&
e^{\mu \ell} 
P\left(\bbox{k},e^\ell,\omega 
e^{\mu \ell},D_2 (\ell),\Delta_L(\ell),\Delta_T (\ell)\right)
\label{scal1}
\\
D_2 (\bbox{k}, \omega, D_2, \Delta_L, \Delta_T) =&&
e^{(2 - \mu)\ell} D_2 \left(\bbox{k} e^\ell,\omega 
e^{\mu \ell}, D_2 (\ell), \Delta_L (\ell), \Delta_T (\ell)\right)~.
\label{scal2}
\end{eqnarray}

For $\epsilon < 0$, i.e., $d > d_{c1}$,. where the stable Gaussian fixed point 
controls the scaling behavior, we obtain, of course, the same scaling 
properties of $P$ as in the isotropic case, i.e., the dynamic exponent 
$z = \mu$ and $P$ is given by the scaling functions in Eqs. (\ref{scal1}) 
and (\ref{scalf}).  For the diffusion coefficient $D_2$ we obtain, 
correspondingly, 

\begin{eqnarray}
D_2 (\bbox{k}, \omega, D_2, \Delta_L, \Delta_T) =
e^{(2 - \mu)\ell} 
D_2\left(\bbox{k} e^\ell,\omega e^{\mu \ell},\Delta_L
e^{-|\epsilon|\ell} , \Delta_T e^{-|\epsilon|\ell}\right)
\end{eqnarray}

In the long wavelength limit $\bbox{k} \rightarrow 0$ and using the 
perturbative result $D_2 \sim \alpha \Delta_L + \beta \Delta_T$ we have, 
setting $\omega e^{\mu \ell} \sim 1$ and $k e^{\mu \ell} \sim 1$, 
the same results as in the isotropic case given by Eqs. (\ref{hom9})
and (\ref{hom10}) and the corresponding results for the temporal and
spatial behavior in Eqs. (\ref{hom11}) and (\ref{hom12}).

Below the critical dimension $d_{c1}$ in the neighborhood of the isotropic 
fixed point in Eq. (\ref{p2}) we obtain from 
Eqs. (\ref{k2}) and (\ref{k3}) the solutions for $\delta \Delta_L$ and 
$\delta \Delta_T$, 

\begin{eqnarray}
\delta \Delta_L (\ell) 
=&& - \frac{(3-d)}{2(d-2)}\left( \delta\Delta_L^0 - \frac{d-1}{3-d}
\delta \Delta_T^0 \right) e^{- |\epsilon|\ell} 
+ \frac{d-1}{2(d-2)} \left( \delta \Delta_L^0-
\delta \Delta_T^0 \right) e^{- |\epsilon| \frac{d-1}{3-d}\ell}\\ 
\delta \Delta_T(\ell) 
=&& - \frac{3-d}{2(d-2)} \left( \delta \Delta_L^0 -\frac{d-1}{3-d}\delta
\Delta_T^0 \right) e^{-|\epsilon|\ell} 
+ \frac{3-d}{2(d-2)} \left( \delta\Delta_L^0-
\delta \Delta^0_T \right) e^{- |\epsilon| \frac{d-1}{3-d}} \ell
\end{eqnarray}
and we conclude that the isotropic fixed pont is stable for $1 < d < d_{c1}$.  

Similarly, near the anisotropic fixed point in Eq. (\ref{p3}) we have

\begin{eqnarray}
 \delta \Delta_L (\ell) =&&
\left( \delta \Delta_L^0 -\frac{d-1}{3-d}\delta\Delta_T^0 
\right) e^{-|\epsilon|\ell} + \frac{d-1}{3-d} \delta \Delta_T^0 
e^{- | \epsilon| \frac{1-d}{3-d}\ell} \\
\delta \Delta_T (\ell) =&& \delta \Delta_T^0 e^{- |\epsilon| \frac{1-d}{2}\ell} 
\end{eqnarray}
and we infer that the anisotropic fixed point is stable for $d < 1$.  
In Fig. 16 we have depicted the renormalization group flow in the 
$(\Delta_T, \Delta_L)$ plane in the 
three cases a) $d> d_{c1}$, b) $d_{c1} >d>1$, 
and c) $d<1$.

Finally, below $d_{c1}$ we obtain for the diffusion coefficient $D_2$ the same 
result as in the isotropic case in Eqs.(\ref{hom5}), (\ref{hom6},
(\ref{hom7}) and (\ref{hom8}).
\subsection{The anisotropic long range case}

We finally discuss the anisotropic long range case.  In Sec. III we have 
given the general renormalization group equations in the case of two 
separate fall-off exponents $a_L$ and $a_T$ for the longitudinal and 
transverse force correlations, respectively.  For simplicity we here only 
consider the case of a common fall-off exponent $a = a_T = a_L < d$.  
Keeping $D_1$ fixed by locking $z$ onto $\mu$ and setting $\Delta_1^{L,T} =0$ 
and $\Delta_2^{L,T} = \Delta_{L,T}$ we extract from 
Eqs. (\ref{rggg1})-(\ref{rggg6}) the renormalization group equations

\begin{eqnarray}
\frac{dD_2}{d\ell} 
=&& (\mu - 2)D_2 
+A\frac{D_1 (2 - \mu - a) - D_2 a}{(D_1 + D_2)^2} \Delta_L 
\nonumber\\
&&+A\frac{D_1 (2d - 2) + D_2 (2 d-2)}{(D_1 + D_2)^2} 
\Delta_T 
\\
\frac{d \Delta_L}{d\ell} 
=&& \epsilon \Delta_L - 
4A\frac{\Delta_L^2}{(D_1 + D_2)^2} 
\\
\frac{d \Delta_T}{d\ell} 
=&& \epsilon \Delta_ - 
4A\frac{\Delta_L \Delta_T}{(D_1 + D_2)^2}~, 
\end{eqnarray}
where we have introduced the expansion parameter
$
\epsilon = 2 \mu - 2 - a = d_{c1} - a~. 
$
To first order in $\epsilon$ we identify the usual Gaussian fixed point
$
(\Delta_L^*, \Delta_T^*) = (0,0)
$
and  a line of non-trivial fixed points
$
\Delta_L^* = \epsilon D_1^2/4 A 
$
.
We note that here there is no isolated non-trivial fixed point unlike in the 
short range case discussed above \cite{fall}.

To linear order in $\epsilon$ we obtain in the vicinity of the respective 
fixed points, setting $\Delta_{L,T} = \Delta_{L,T}^* + \delta \Delta_{L,T}$,

\begin{eqnarray}
\frac{ d \delta \Delta_L}{d\ell} 
&=& \left( \epsilon - \frac{8A}{D_1^2} \Delta_L^* \right) 
\delta \Delta_L 
\\
\frac{d \delta \Delta_T}{d\ell} 
&=& \left( \epsilon - \frac{4 A}{D_1^2} \Delta_L^* \right) \delta  
\Delta_T - \frac{4 A}{D_1^2} \Delta_T^* \delta \Delta_L ~.
\end{eqnarray}
In the neighborhood of the Gaussian fixed point we have the solutions
$
\delta \Delta_T (\ell) 
= \delta \Delta^0_T e^{\epsilon \ell} 
$
and
$
\delta \Delta_L (\ell) 
= \delta \Delta_L^0 e^{\epsilon \ell}~,
$
showing that the fixed point is stable for $\epsilon < 0$, i.e., for 
$d > a> 2 \mu - 2$ (note that $a < d$ in the long range case).  
Eliminating the scaling parameter $\ell$ we obtain for the flow in the 
$\Delta_L - \Delta_T$ plane
$
\delta \Delta_T=(\delta\Delta_T^0/\delta\Delta_L^0)\delta \Delta_L 
$
,
i.e., the trajectories approach the fixed point with constant slope 
\cite{fall}.  Similarly for the scale dependent diffusion coefficient 
$D_2 (\ell)$ we have 

\begin{eqnarray}
\delta D_2 =
\left[ \delta D_2^0 + \frac{A}{D_1} 
\left(\frac{2 - \mu - a}{d - \mu} \delta 
\Delta_L^0 + \frac{d (d-1)}{a - \mu} \delta 
\Delta_T^0 \right) \right] e^{(2-\mu)\ell}
- \frac{A}{D_1} \left[ \frac{2 - \mu - a}{a - \mu} 
\delta \Delta_L^0 + \frac{2 (d-1)}{a - \mu} 
\delta \Delta_T^0 \right] e^{\epsilon \ell}~.
\end{eqnarray}
Eliminating $\ell$ we obtain for the flow for $\epsilon < 0$

\begin{eqnarray}
\delta D_2 =&&\left[\delta D_2^0+\frac{A}{D_1} 
\left(\frac{2-\mu-a}{a-\mu}\delta 
\Delta_L^0+\frac{2(d-1)}{a-\mu}\delta 
\Delta_T^0\right)\right]\left[\frac{\delta 
\Delta_{L,T}}{\delta\Delta_{L,T}^0}\right]^{(2 - \mu)/| \epsilon|} 
\nonumber \\
&&-\frac{A}{D_1}\left[\frac{2-\mu-d}{a-\mu} 
\delta \Delta_L+\frac{2(d-1)}{a-\mu}\delta\Delta_T\right]
\end{eqnarray}
and the discussion in the anisotropic short range case
applies with $d$ replaced by $a$.

The scaling analysis of $P$ and $D_2$ also proceeds as before.  We have the 
general homogeneity relations in Eqs. (\ref{scal1}-\ref{scal2}).
For $\epsilon < 0$, i.e., $a > 2 \mu - 2$, the Gaussian fixed point controls 
the scaling behavior.  The dynamic exponent $z = \mu$ and $P$ is given by 
Eqs. (\ref{scal1}) and (\ref{scalf}).  For $D_2$ we obtain

\begin{eqnarray}
D_2 (\bbox{k}, \omega, D_2, \Delta_L, \Delta_T) =
e^{(2 - \mu)\ell} D_2 \left(\bbox{k} e^\ell, \omega 
e^{\mu \ell}, \Delta_L e^{- |\epsilon| \ell}, 
\Delta_T e^{-|\epsilon|\ell} \right)
\end{eqnarray}
and using the perturbative result 
$D_2 \propto\alpha \Delta_L + \beta \Delta_T$ we find, 
setting $\omega e^{\mu\ell} \simeq 1,$
and $\bbox{k}e^{\ell}\simeq 1$, respectively, 

\begin{eqnarray}
D_2(\bbox{0},\omega,D_2,\Delta_L,\Delta_T)\propto&&\omega^{(a-\mu)/\mu}\\
D_2(\bbox{k},0,D_2,\Delta_L,\Delta_T)\propto&& k^{a-\mu}
\end{eqnarray}
and
\begin{eqnarray}
D_2 (\bbox{0}, t, D_2, \Delta_L, \Delta_T) 
&\propto& t^{- a/\mu} 
\\
D_2 (\bbox{r}, 0, D_2, \Delta_L, \Delta_T) 
&\propto& r^{\mu - d - a}
\end{eqnarray}
For $\epsilon > 0$, i.e., $a < 2 \mu - 2$, we have in the vicinity 
of the stable 
line of fixed points
$
\delta \Delta_L (\ell) 
=\delta \Delta_L^0 e^{- \epsilon \ell} 
$
and 
$
\delta \Delta_T (\ell) 
=  \delta \Delta_T^0 + \delta \Delta_L^0 
(\Delta_T^*/\Delta_L^*)(e^{- \epsilon \ell} - 1)
$
or eliminating $\ell$
$
\delta \Delta_T - \delta \Delta_T^0 = 
(\Delta_T^*/\Delta_L^*)(\delta \Delta_L - \delta \Delta_L^0)
$
,
showing that the trajectories approach the line of fixed points 
$\Delta_L^* = \epsilon_L D_1^2 / 4 A$ with slope $\Delta_T^*/\Delta_L^*$.
In Fig. 17 we have shown the renormalization group flow in the 
$\Delta_T - \Delta_L$ plane in the cases a)  $a > d_{c1}$ and b) $a< d_{c1}$.
\draft

\centerline{\bf\large Figure captions}
\smallskip
\smallskip
\smallskip
\noindent
Fig. 1. Plot of the L\'{e}vy distribution for a microscopic step
$\eta$.  The tail of $p(\bbox{\eta})$ is characterized by the step 
index $f$.  The
distribution is cut off at a smallest step distance $\eta_0$ corresponding
to a microscopic length.
\smallskip

\noindent
Fig. 2. The scaling index $\mu$ plotted as a function of the step
index $f$.
\smallskip

\noindent
Fig. 3. Diagrammatic representation of the 
Fokker-Planck equation.
\smallskip

\noindent
Fig. 4. Diagrammatic representation of 
the Dyson equation defining the self energy 
$\Sigma(\bbox{k},\omega)$ and the renormalized 
propagator $G(\bbox{k},\omega)$.
\smallskip

\noindent
Fig. 5. Diagrammatic representation of 
the renormalized Fokker-Planck equation.
\smallskip

\noindent
Fig. 6. Diagrammatic representation of the 
4-point vertex function 
$\Gamma (\bbox{k}, \bbox{p}, \bbox{l}, \omega)$ 
and the contraction 
$\int\Gamma(\bbox{k},\bbox{k},\bbox{l},\omega)
G(\bbox{k}-\bbox{l},\omega)d^d\ell/(2\pi)^d$.
\smallskip

\noindent
Fig. 7. Diagrammatic representation of the 
first loop order correction to $\Sigma(\bbox{k},\omega)$.
\smallskip

\noindent
Fig. 8. Diagrammatic representation of the 
first loop order correction to ${\bf \Lambda} (\bbox{k}, \bbox{p}, \omega)$.
\smallskip

\noindent
Fig. 9. Diagrammatic representation of the 
first loop order corrections 
to $\bbox{\Gamma}(\bbox{k},\bbox{p},\bbox{l},\omega)$.
\smallskip

\noindent
Fig. 10. Diagrammatic representation of the renormalized 
Fokker-Planck equation valid in the long wavelength 
region $1< k < e^{-\ell}$.  The slash indicates corrections 
evaluated on the shell $e^{-\ell} < k < 1$.
\smallskip

\noindent
Fig. 11. Plot of the critical dimension $d_{c1}$ as 
a function of the scaling index $\mu$.  For $\mu = 2$ we 
have the Brownian case $d_{c1} = 2$; for $1 < \mu < 2$ 
the dimension $d_{c1}$ depends linearly on $\mu$.  
The ballistic case $\mu = 1$ is attained for $d_{c1} = 0$.  
The critical dimension $d_{c1}=2\mu-2$ and the line $d = \mu$
separate the regions I, II and II, III, respectively.
In I the diffusion coefficient $D_2$ vanishes
and the disorder is {\it irrelevant},
in II $D_2$ diverges and disorder is {\it irrelevant},
and in III $D_2$ diverges and disorder becomes {\it relevant}.
\smallskip

\noindent
Fig. 12. Plot of the fixed point 
$(\Delta^*, D_2^*)$ as a function of the scaling 
index $\mu$ for $1 + d/2 < \mu < 2$.
\smallskip

\noindent
Fig. 13. Plot of the fall-off exponent 
as a function of the scaling index $\mu$.  For 
$\mu = 2$ we have the Brownian case.  The lines 
$a = \mu$ and $a = 2 \mu - 2$ separate the 
regions I, II and II, III, respectively.  
In I the diffusion coefficient $D_2$ vanishes 
and the disorder is {\it irrelevant}, 
in II $D_2$ diverges and disorder is {\it irrelevant}, 
and in III $D_2$ diverges and disorder becomes {\it relevant}.
\smallskip

\noindent
Fig. 14. Plot of $a$ versus $d$.  The line $a = d$ 
delimits the short range and long range regions.  
In region III for $a > \mu$ and $d > \mu$ the 
diffusion coefficient $D_2$ vanishes and the disorder 
is {\it irrelevant}, in region II for 
$\mu > d > 2 \mu - 2$ and $\mu > a > 2 \mu - 2$ ~$D_2$ 
diverges and disorder is {\it irrelevant}, and 
in region III for $0 < d < \mu$ and $ 0 < a < \mu$ ~$D_2$ 
diverges and disorder becomes {\it relevant}.
\smallskip

\noindent
Fig. 15. Renormalization group flow in
the $\Delta - D_2$ plane about the Gaussian
fixed point $(\Delta^*, D_2^*) = (0,0)$.
In case a) $d > \mu$, case b) $d = \mu$, and
case c) $d_{c1} < d< \mu$.
\noindent
Fig. 16. {\bf The anisotropic short range case}:
Renormalization group flow in the 
$\Delta_T - \Delta_L$ plane.  Case a): $d > d_{c1} = 2 \mu - 2$ 
and the trajectories flow towards the stable Gaussian 
fixed point $(G)$ with constant slope.  
Case b):  $1 < d < d_{c1}$ and the trajectories 
flow towards the non-trivial isotropic fixed point 
$(I)$.  Case c):  $d < 1$ and the trajectories flow 
towards the non-trivial longitudinal fixed point $(A)$.
\smallskip

\noindent
Fig. 17. {\bf The anisotropic long range case}:
Renormalization group flow in the 
$\Delta_T - \Delta_L$ plane.  Case a):  
$a> d_{c1} = 2 \mu - 2$ and the trajectories flow towards 
the stable Gaussian fixed point $(G)$ with constant slope.  
Case b):  $a < d_{c1} = 2 \mu - 2$ and the trajectories 
flow towards the non-trivial line of fixed points 
$(L)$ with constant slope.

\begin{references}
\bibitem{fall}
We note that introducing distinct fall-off exponents 
$a_L \neq a_T$, i.e., $\epsilon_L \neq \epsilon_T$, 
the ``degeneracy'' is lifted and instead of a line of 
fixed points we only obtain the longitudinal fixed point
\end{references}
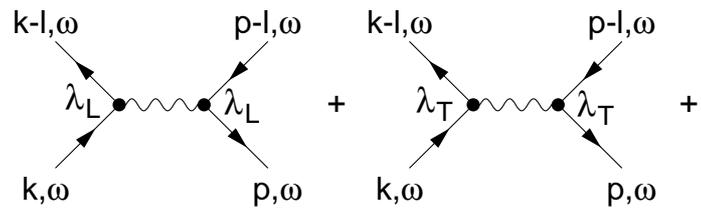
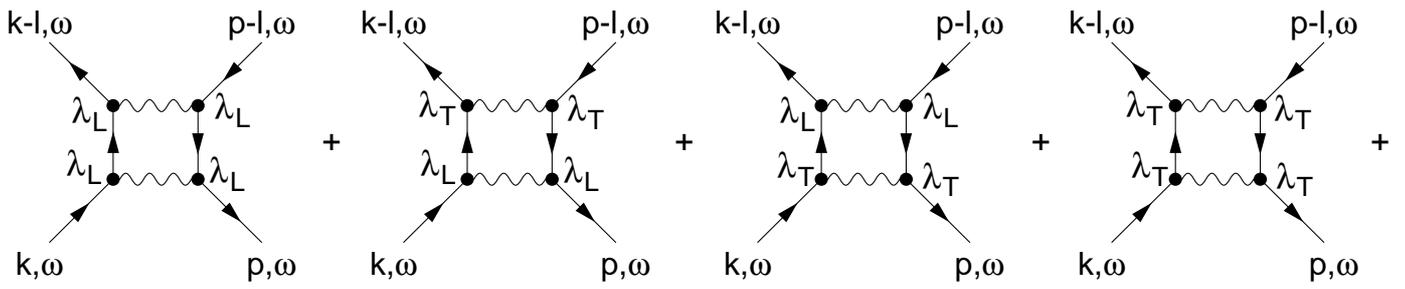
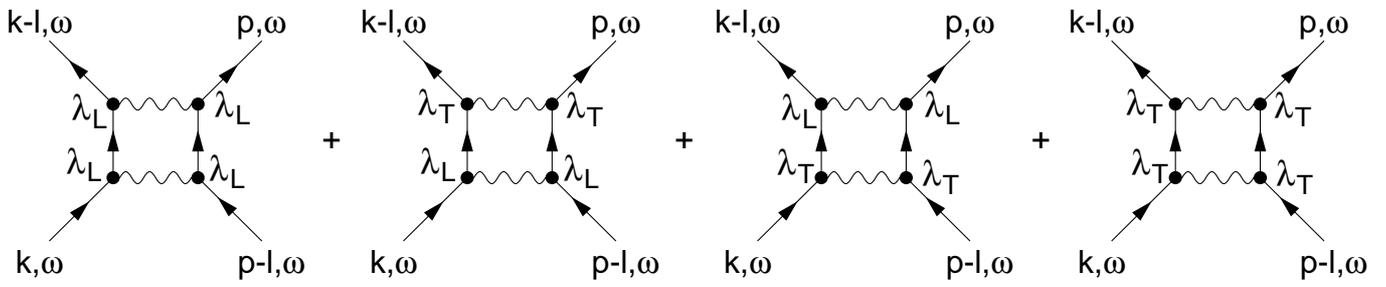
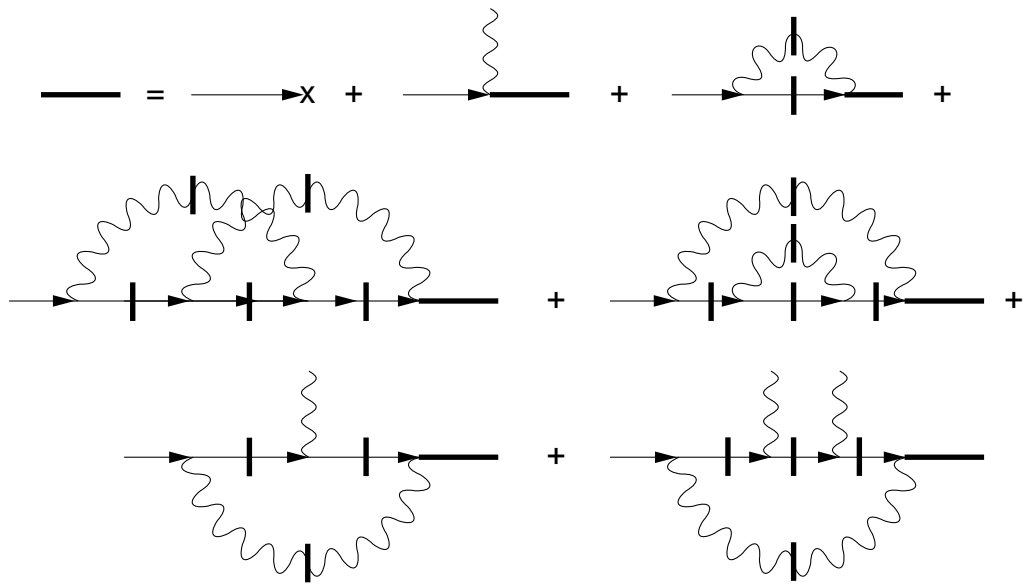
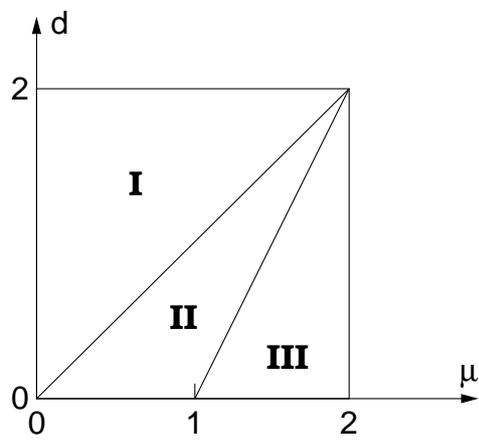
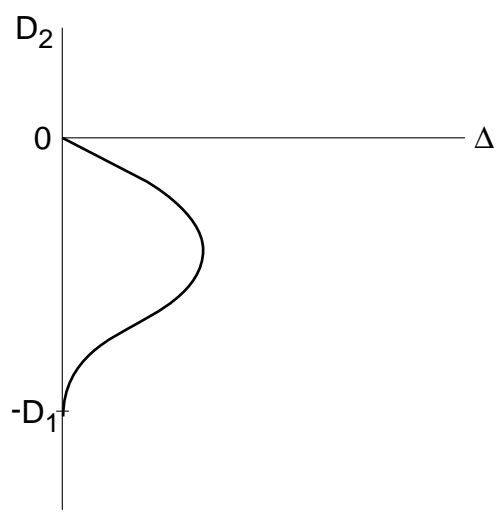
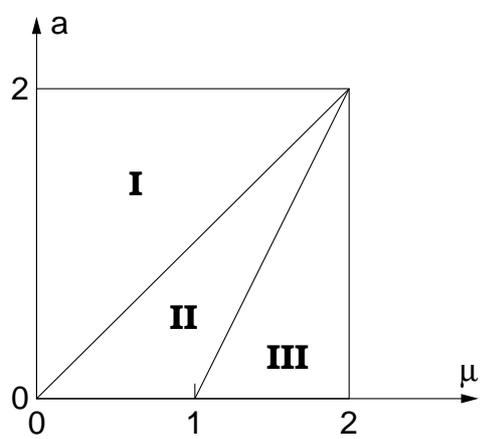
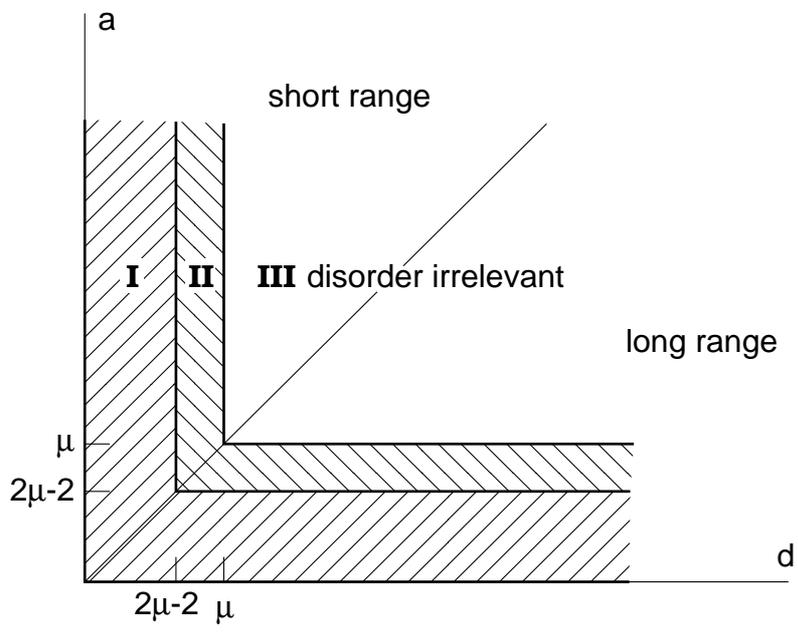
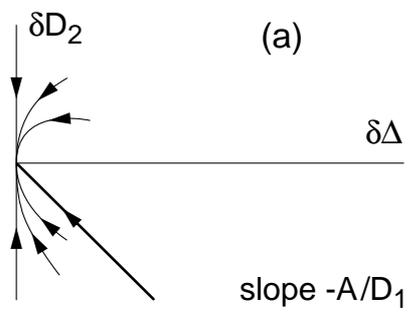
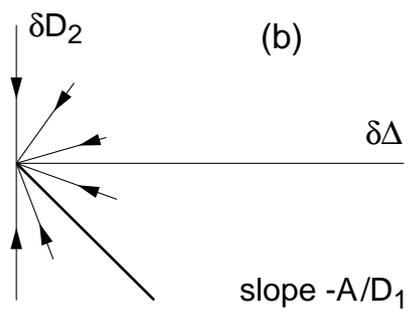
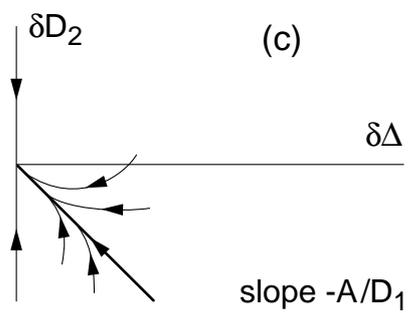
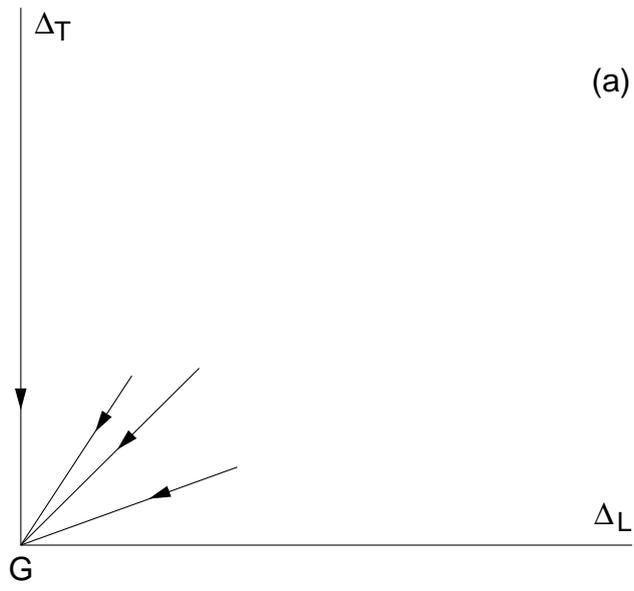
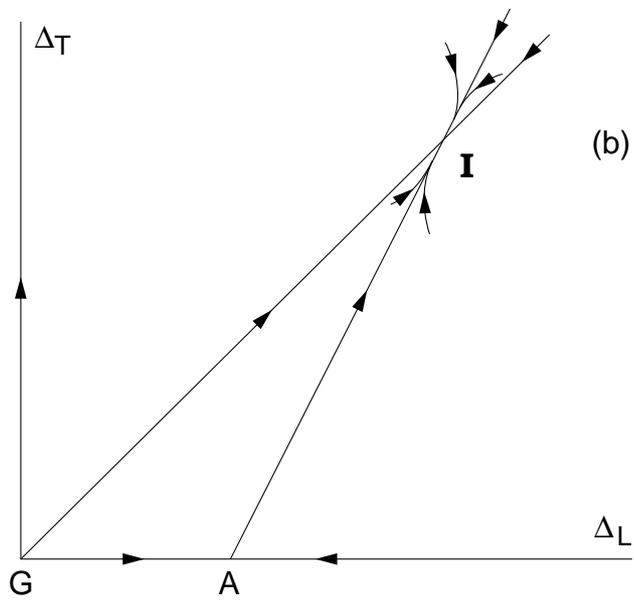
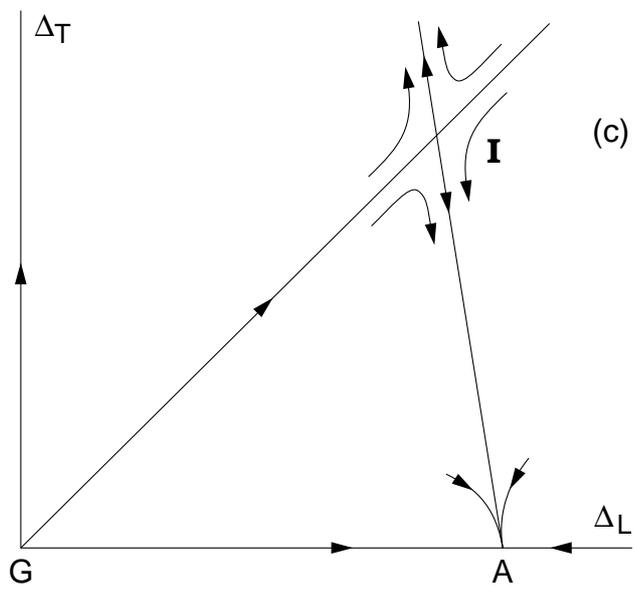
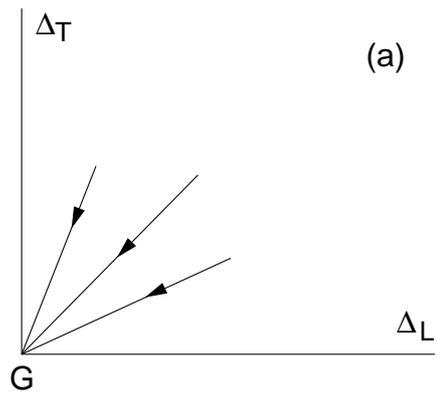
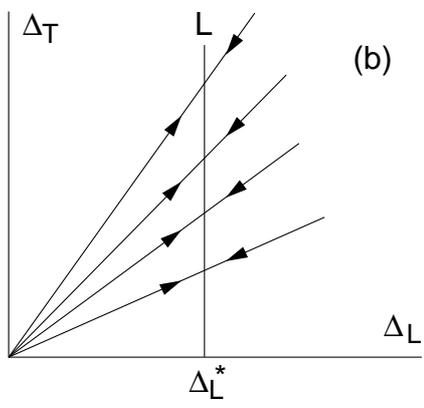
\end{document}